\documentclass[preprint,superscriptaddress,nofootinbib,preprintnumbers,amsmath,amssymb,notitlepage,11pt]{revtex4-1}
\usepackage[T1]{fontenc}

\usepackage[titles]{tocloft}
\cftsetindents{section}{0em}{2.5em}
\cftsetindents{subsection}{2.5em}{2.5em}

\usepackage{amsfonts}
\usepackage{mathrsfs}
\usepackage{leftidx}
\usepackage{amssymb}
\usepackage{placeins}
\usepackage{relsize}
\usepackage{slashed}

\usepackage[usenames,dvipsnames,svgnames]{xcolor}
\definecolor{navy}{rgb}{0.05, 0.23,0.85}

\usepackage[colorlinks]{hyperref}
\hypersetup{
     colorlinks   = true,
     citecolor    = red,
	linkcolor = navy
}

\usepackage{soul}
\usepackage{epsfig}
\usepackage{graphicx}               

\usepackage{url}
\usepackage{float}
\usepackage{color}

\newcommand{\be}{\begin{equation}}
\newcommand{\ee}{\end{equation}}
\newcommand{\bea}{\begin{eqnarray}}
\newcommand{\eea}{\end{eqnarray}}
\newcommand{\beq}{\begin{eqnarray}}
\newcommand{\eeq}{\end{eqnarray}}

\newcommand{\eV}{\textrm{eV}}
\newcommand{\keV}{\textrm{keV}}
\newcommand{\MeV}{\textrm{MeV}}
\newcommand{\GeV}{\textrm{GeV}}

\newcommand{\Neff}{N_\textrm{eff}}

\graphicspath{{./figures/}}

\usepackage{enumitem}
\newlist{steps}{enumerate}{1}
\setlist[steps, 1]{label = Step \arabic*:}

\begin{document}


\title{
Light Dark Matter: Models and Constraints
}
\author{Simon Knapen}
\affiliation{Theory Group, Lawrence Berkeley National Laboratory, Berkeley, CA 94709 USA}
\affiliation{Berkeley Center for Theoretical Physics, University of California, Berkeley, CA 94709 USA}
\author{Tongyan Lin}
\affiliation{Theory Group, Lawrence Berkeley National Laboratory, Berkeley, CA 94709 USA}
\affiliation{Berkeley Center for Theoretical Physics, University of California, Berkeley, CA 94709 USA}
\affiliation{Department of Physics, University of California, San Diego, CA 92093, USA }
\author{Kathryn M. Zurek}
\affiliation{Theory Group, Lawrence Berkeley National Laboratory, Berkeley, CA 94709 USA}
\affiliation{Berkeley Center for Theoretical Physics, University of California, Berkeley, CA 94709 USA}

\begin{abstract}
We study the direct detection prospects for a representative set of simplified models of sub-GeV dark matter (DM), accounting for existing terrestrial, astrophysical and cosmological constraints. 
We focus on dark matter lighter than an MeV, where these constraints are most stringent, and find three scenarios with accessible direct detection cross sections:  {\em (i)} DM interacting via an ultralight kinetically mixed dark photon,  {\em (ii)}  a DM sub-component interacting with nucleons or electrons through a light scalar or vector mediator, and  {\em (iii)} DM coupled with nucleons via a mediator heavier than $\sim 100 \mbox{ keV}$.
\end{abstract}

\date\today

%
%
%
\maketitle
\flushbottom
%
\clearpage
\setcounter{tocdepth}{2}
\tableofcontents


\section{Introduction}

In recent years, new ideas to search for dark matter (DM) have changed the direct detection landscape.  As highly sensitive searches for the Weakly Interacting Massive Particle (such as LUX~\cite{Akerib:2016vxi}, PandaX-II~\cite{Cui:2017nnn}, XENON1T~\cite{Aprile:2017iyp} and SuperCDMS~\cite{Agnese:2017njq}) have not yet turned up a signal, the urgency to develop techniques to search for DM outside of the $\sim 10 \mbox{ GeV} - 10 \mbox{ TeV}$ mass range has increased.  Outside of this window, there are theoretically well-motivated candidates that are consistent with the observed history of our Universe.  DM may reside in a hidden sector communicating via a mediator coupled to both sectors (see for example \cite{Boehm:2003hm,Strassler:2006im,Hooper:2008im,Pospelov:2007mp,Feng:2008ya,Morrissey:2009ur}). The relic abundance can be fixed in a variety of ways, including via particle-anti-particle asymmetry as in asymmetric DM~\cite{Kaplan:2009ag,Petraki:2013wwa,Zurek:2013wia}, strong dynamics~\cite{Boddy:2014yra,Hochberg:2014dra}, freeze-in~\cite{Hall:2009bx,Bernal:2017kxu}, or other thermal histories for example \cite{Kuflik:2015isi,DAgnolo:2017dbv,Berlin:2017ftj,Pappadopulo:2016pkp,DAgnolo:2015ujb,Dror:2016rxc,Heeck:2017xbu}.   
Cosmology (via structure formation bounds on warm DM) indicates that such DM candidates, populated through thermal contact with the standard model (SM) at some epoch, must be heavier than $\sim \mbox{few keV}$ to tens of keV~\cite{Yeche:2017upn,Irsic:2017ixq}. Thus the $\mbox{ keV} - 10 \mbox{ GeV}$ mass window is a natural place to consider searching for light DM.

Because of the existence of well-motivated candidates (such as asymmetric DM) pointing to the GeV scale, a number of direct detection experiments targeting WIMPs through nuclear recoils have also been actively working to increase their sensitivity to smaller energy depositions and lighter DM candidates. Such experiments include CRESST-II~\cite{Angloher:2015ewa}, DAMIC~\cite{Barreto:2011zu}, NEWS-G~\cite{Arnaud:2017bjh}, PICO~\cite{Amole:2017dex}, SENSEI~\cite{Tiffenberg:2017aac}, and SuperCDMS~\cite{Agnese:2013jaa,Agnese:2015nto,Agnese:2016cpb,Agnese:2017jvy}. In addition, there are a number of recent proposals to detect nuclear recoils of sub-GeV dark matter, such as liquid helium detectors~\cite{Guo:2013dt}, bond breaking in molecules~\cite{Essig:2016crl}, and defect creation in crystal lattices~\cite{Kadribasic:2017obi,Budnik:2017sbu}.  
Furthermore, for DM masses in the MeV-GeV range, the largest energy depositions are typically achieved not via nucleon interactions, but via scattering off electrons.  New detection methods sensitive to electron interactions have thus been proposed with semiconductors~\cite{Essig:2011nj,Essig:2015cda}, atoms~\cite{Essig:2012yx}, graphene~\cite{Hochberg:2016ntt}, and scintillators~\cite{Derenzo:2016fse}.

When the DM mass is below a MeV, new targets and detection techniques sensitive to meV energy depositions must be sought.  Superconductors~\cite{Hochberg:2015pha,Hochberg:2015fth,Hochberg:2016ajh} and Dirac materials~\cite{Hochberg:2017wce} have been proposed to detect sub-MeV DM with coupling to electrons, while superfluid helium~\cite{Schutz:2016tid,Knapen:2016cue}, though multi-phonon and multi-roton excitation, can have sensitivity to such light DM with coupling to nucleons. The same experiments are sensitive to small energy depositions can also detect bosonic DM (produced nonthermally) via absorption \cite{Hochberg:2016ajh,Hochberg:2016sqx,Bloch:2016sjj,Bunting:2017net,Arvanitaki:2017nhi}.  (See Ref.~\cite{Battaglieri:2017aum} for a recent summary of dark matter detection proposals.)

Direct detection experiments do not, however, exist in isolation from other types of probes.  If the DM particle can be probed via electron or nucleon couplings in direct detection experiments, this necessarily implies the presence of other constraints from the early Universe, structure formation in the late Universe, as well as laboratory probes. Most of these constraints are model dependent in one way or another, and to get a sense of their relation to the size of the couplings probed in direct detection experiments, one must commit to a set of benchmark models. Our goal in this work is to present a unified and complete picture of these constraints on a variety of simplified models for sub-GeV DM, for interactions with both electrons and nucleons.  For DM having mass between an MeV and a GeV, the importance of cosmological history and terrestrial constraints was highlighted by for instance in Refs.~\cite{Krnjaic:2015mbs,Boehm:2013jpa,Batell:2014mga,Essig:2013vha}. When the DM becomes lighter than an MeV, a tapestry of DM self-interaction, stellar emission, $\Neff$ and terrestrial constraints comes to the fore.  When the DM has an electron coupling, some of these constraints were discussed in Ref.~\cite{Hochberg:2015pha} and considered more extensively in Ref.~\cite{Hochberg:2015fth,Green:2017ybv}, while scalar-mediated nucleon couplings have also been discussed in Ref.~\cite{Green:2017ybv}.

In all our benchmark models, the DM interacts with the SM via the exchange of a mediator (whether scalar or vector) in the $t$-channel. The mediator will be constrained by virtue of the fact that it couples to SM particles. The scattering cross section in direct detection experiments is typically parametrized by
\beq
	\bar \sigma_{\rm DD} \sim \frac{4 \pi \alpha_T \alpha_\chi}{(m_\phi^2 + {\bf q}_0^2)^2} \mu_{T\chi}^2,
\eeq
where $\mu_{T\chi}$ is the reduced mass of the DM $\chi$ with the target (whether electron or nucleon), and ${\bf q}_0$ is a reference value for the momentum transfer. (Precise equations and conventions will be established for each benchmark model in the corresponding section.) The scattering is defined by the light or heavy mediator regimes, when the momentum transfer $q$ is much smaller or much larger than the mediator mass $m_\phi$.  In these regimes, the scattering cross section is
\bea
\label{eq:DDCrossSections}
\sigma_{\rm DD}^{\rm massless} & \simeq & 1 \times 10^{-39} \mbox{ cm}^2 \left(\frac{\alpha_T \alpha_\chi}{10^{-30}}\right)\left(\frac{\mu_{T\chi}}{m_e}\right)^2 \left(\frac{\mbox{keV}}{q}\right)^4 \\ \nonumber
\sigma_{\rm DD}^{\rm massive} & \simeq & 2 \times 10^{-40} \mbox{ cm}^2 \left(\frac{\alpha_T \alpha_\chi}{10^{-16}}\right)\left(\frac{\mu_{T\chi}}{m_e}\right)^2 \left(\frac{5 \mbox{ MeV}}{m_\phi}\right)^4.
\eea
For light mediators, even with small couplings the rate is potentially observable in a direct detection experiment sensitive to low momentum-transfer scattering.  Clearly, to understand the parameter space for direct detection, we must understand the nature of the constraints on the couplings of the mediator to the DM, $\alpha_\chi$, and of the mediator to the target, $\alpha_{T}$.  The relevant constraints depend of course on the mass and spin of the mediator, and whether the couplings are predominantly to nucleons, electrons or both. The interplay between the various bounds is best understood in terms of the mediator mass regime in which they dominate.  

{\em \underline{Massive mediator}}:  In the massive mediator regime ($m_\phi \gtrsim 1$ MeV),  stellar constraints are absent or substantially reduced such that the couplings of the mediator to the target can be as large as $\alpha_T \sim 10^{-9}$.  The remaining bounds on these couplings primarily come from rare meson processes (such as $B \rightarrow K \phi$) and beam dump experiments. DM self-interaction bounds place a mild limit on $\alpha_\chi$, 
\begin{align}
	\alpha_\chi \lesssim  0.02 \left(\frac{1 \mbox{ keV}}{m_\chi}\right)^{1/2} \left(\frac{m_\phi}{1 \mbox{ MeV}}\right)^2.
\end{align}
For realistic direct detection cross sections, $\alpha_\chi$ and $\alpha_T$ are however large enough that the mediator and the DM are generally in thermal equilibrium with the SM in the early universe. This can lead to a contribution to the number of relativistic degrees of freedom (parameterized in terms of number of effective neutrino species $N_{\rm eff}$), with constraints from Big Bang Nucleosynthesis (BBN) and the Cosmic Microwave Background (CMB).

{\em \underline{Massless mediator}}: When the mediator mass is below $m_e$, stellar constraints generally put a very strong bound on $\alpha_T$, such that only $m_\phi\ll 1$ MeV gives a realistic direct detection cross section. For example, when $m_\phi < 100 \mbox{ keV}$, the constraints require $\alpha_n \lesssim 8\times10^{-26}$ for scalar nucleon couplings.   Stellar constraints in the massless regime are, however, strongly model dependent -- a kinetically mixed dark photon, for example, has a much smaller production in the star than a scalar. For a light mediator, DM self-interactions also become important; for example, the constraint on the coupling $\alpha_\chi$ is:
\begin{align}
	\alpha_\chi \lesssim 6\times 10^{-10} \times  \left(\frac{m_\chi}{1 \mbox{ MeV}}\right)^{3/2}
\end{align}
for $m_\chi v_{DM} / m_\phi\approx 10$ and $v_{DM}\approx 10^{-3}$. However, this constraint is much weaker if we consider a relic $\chi$ which is only a sub-component of all the DM.
  
Here, we study three broad classes of simplified models: 
\begin{itemize}
\item  {\emph{ Real scalar dark matter coupled to nucleons through a hadrophilic scalar. }} A scalar mediator interacting with nucleons can be generated by the scalar coupling to top quarks or to heavy colored vector-like fermions. We show the corresponding constraints in Fig.~\ref{fig:mphiplane}. Models of this type also generate mediator-pion interactions, which are not relevant for direct detection but do matter for the thermal history of the universe. We find, generally, that there are two parameter regimes where sub-MeV DM may be detectable in a low threshold experiment ({\emph{e.g.}} a superfluid helium target \cite{Schutz:2016tid,Knapen:2016cue}) with a kg-year exposure. 

In the first case, the DM scatters via a very light mediator (typically keV mass or lighter) having small enough couplings such that it does not cool stars.  For such small couplings the mediator also decouples from the SM thermal bath before the QCD phase transition, leading to a contribution of the mediator to $\Delta \Neff$ of at most:
\begin{align}
	\Delta \Neff\approx  \frac{4}{7} \left( \frac{g_{SM}(T_{\nu dec})}{g_{SM}(T_{QCD})} \right)^{4/3} \approx 0.06 
\end{align}
with $g_{SM}(T_{\nu dec})$ and $g_{SM}(T_{QCD})$ the number of degrees of freedom in the SM thermal bath at neutrino decoupling and before the QCD phase transition, respectively. Note that this value of $\Delta N_{\rm eff}$ can be tested by CMB S4 experiments \cite{Abazajian:2016yjj}. In order to have large enough direct detection cross sections, we consider a sub-component of the DM so as to evade constraints from DM self-interactions, as can be seen in Fig.~\ref{fig:scenA_money}.

The second case is where the mediator is fairly massive (typically in the 100 keV to 1 MeV mass range). Here, for detectable cross sections, the coupling is large enough that the mediator tends to thermalize with the pions in the early universe, giving rise to $\Delta \Neff \approx 4/7$ for a sub-MeV mediator. This value is in tension at the 2$\sigma$ level with the $\Neff$ derived from recently improved measurements of the deuterium abundance~\cite{Cooke:2013cba,Cyburt:2015mya}. The scenario where the mediator and the dark matter \emph{both} thermalize with the SM is moreover firmly excluded, and we must place a strong limit on $\alpha_\chi$ to avoid thermalization of the DM with the mediator. The resulting parameter space is shown in Fig.~\ref{fig:scenB_money}. 

\item  {\emph{ Real scalar dark matter coupled to an leptophilic scalar mediator. }}   Similar to the nucleon case, a mediator coupling to the electrons is constrained by fifth force experiments and stellar cooling arguments when the mediator is light, and predominantly by beam dump and other accelerator experiments when the mediator is heavier. Bounds are shown in Fig.~\ref{fig:mphiplane_electron}. (As such, this model shares many qualitative features with the Higgs-portal dark matter model considered in Ref.~\cite{Krnjaic:2015mbs} for $m_{\rm DM}>1$ MeV.)  Similar to the nucleon case, for sub-MeV DM scattering via a light mediator we find sizable direct detection cross sections only for a sub-component of the total DM; this is illustrated in Fig.~\ref{fig:scenA_money_phiee}. For the massive mediator case, BBN constraints from thermalization of the mediator via its couplings to the electron are particularly strong due to improved deuterium measurements~\cite{Cyburt:2015mya}.  As shown in Fig.~\ref{fig:scenB_money_phiee}, this thermalization consideration implies that a low-threshold experiment (such as a superconducting detector with kg-year exposure \cite{Hochberg:2015pha,Hochberg:2015fth}) will not have sensitivity to sub-MeV DM scattering via a massive mediator. 

\item  {\emph{ Dirac fermion dark matter, coupled to a kinetically mixed dark photon or a $B-L$ gauge boson.}}  Because of strong constraints from BBN, we consider only the case of a light mediator with sufficiently small couplings that it does not thermalize with the SM.  As has been noted elsewhere \cite{Essig:2015cda,Hochberg:2017wce}, and shown explicitly in Fig.~\ref{fig:darkphoton}, scattering via a kinetically mixed dark photon is consistent with all current bounds in the 10 keV - 1 GeV mass range.  For $m_\chi\lesssim 1$ MeV, such DM can be probed with Dirac materials and superconductors, although in the latter case the reach is substantially reduced due to in-medium effects. For $m_\chi>1$ MeV various other targets have sensitivity, in particular semiconductors. For the $B-L$ gauge boson, there are strong fifth force and stellar constraints, as can be seen in Fig.~\ref{fig:BL_crosssection}.  Scattering of sub-MeV DM through such a mediator would be detectable by either a superfluid helium or a Dirac material target, but only for sub-component DM that evades self-interaction constraints, shown in Fig.~\ref{fig:BL_crosssection}.

\end{itemize}
This paper is organized to consider each of these models in turn: scalar DM coupling to a hadrophilic scalar mediator in Section~\ref{sec:scalarnucleon}, to a leptophilic scalar mediator in Section~\ref{sec:scalarelectron}, and DM scattering via a vector mediator in Section~\ref{sec:otherscenarios}.  We summarize the results in Section~\ref{sec:conclusion}.  We emphasize that current prospects are often dictated by the capabilities of particular target materials for detection of sub-MeV dark matter -- in the case of nucleon couplings, superfluid helium  and in the case of electron couplings, superconductors and Dirac materials.  The general considerations studied in this paper motivate the search for materials with even stronger sensitivity to light dark matter.

\section{Hadrophilic scalar mediator \label{sec:scalarnucleon}}
In our first model we assume a hadrophilic scalar mediator $\phi$ has couplings exclusively to the SM hadrons, specifically pions and nucleons. The coupling to nucleons is the most consequential for direct detection and the majority of the constraints, and we therefore parametrize the model in terms of the low energy effective Lagrangian
\begin{align}
	\label{eq:nucleonmodel}
	\mathcal{L}\supset - \frac{1}{2}m_\chi^2\chi^2 - \frac{1}{2}m_\phi^2 \phi^2 - \frac{1}{2}y_\chi m_\chi \phi \chi^2 - y_n \phi \overline n n, 
\end{align}
where $\chi$ is a real scalar that composes all or part of the DM.
On its own, this potential has runaway directions, but these can be stabilized by adding in quartic couplings for the scalars without affecting the dark matter phenomenology.  One may further verify that with the normalization in Eq.~\eqref{eq:nucleonmodel}, the perturbativity condition is $y_\chi\lesssim 4\pi$ for $m_\phi\ll m_\chi$, although we will conservatively require $y_\chi<1$.  This will be relevant for the light mediator regime, where we will consider sub-component dark matter such that the self-interaction constraints are relaxed. We further elaborate on unitarity and vacuum stability in Appendix~\ref{app:stability}.

In order to account for both the heavy and light mediator limits, we will parametrize the direct detection cross section on nucleons as
\begin{equation}
	\sigma_n \equiv \frac{y_n^2 y_\chi^2}{4\pi}\frac{\mu_{\chi n}^2}{(m_\phi^2 + v_{DM}^2 m_\chi^2)^2},
\end{equation}
where $\mu_{\chi n}$ is the DM-nucleon reduced mass, and $m_\chi v_{DM}$ is a reference momentum transfer with $v_{DM}=10^{-3}$. The dark matter scattering ``form factors'' (as defined, for example, in Ref.~\cite{Essig:2011nj}) in the heavy and light mediator limits are  $F^2(q^2)=1$ and $F^2(q^2)=(v_{DM} m_\chi)^4/ q^4$, respectively. With the normalization of the Yukawa coupling in Eq.~\eqref{eq:nucleonmodel}, the direct detection cross section also has the same scaling as for fermionic $\chi$. The main difference in our results comes in when we consider the cosmology and the contributions of the dark sector to $\Neff$. As we will see, both the scalar and the fermionic DM case are in tension with BBN measurements if they are in equilibrium with the SM below the QCD phase transition. If the dark sector decouples \emph{before} the QCD phase transition, the resulting $\Neff$ depends on the number of degrees of freedom of $\chi$ and could be observable with CMB Stage IV.

Below we outline the primary constraints on $y_n$ and $y_\chi$ arising from meson decays, fifth-force experiments, stellar emission, and DM self-interactions. We then turn to constraints from cosmology, which are sensitive to  thermalization of the mediator and/or the DM. For $m_\chi$ between 100 MeV and 1 GeV, this model can be probed in direct detection experiments such as CRESST~\cite{Angloher:2015ewa}, SuperCDMS~\cite{Agnese:2016cpb} and NEWS~\cite{Profumo:2015oya,Arnaud:2017bjh}. 
In addition, for $m_\chi \ll 1$ GeV, this model could be accessible to proposed low-threshold experiments with superfluid helium \cite{Guo:2013dt,Carter:2016wid,Schutz:2016tid,Knapen:2016cue}, crystal defect techniques \cite{Essig:2016crl,Budnik:2017sbu,Rajendran:2017ynw} and magnetic bubble chambers \cite{Bunting:2017net}. 

\begin{figure}
\centering
\includegraphics[width=0.75\textwidth]{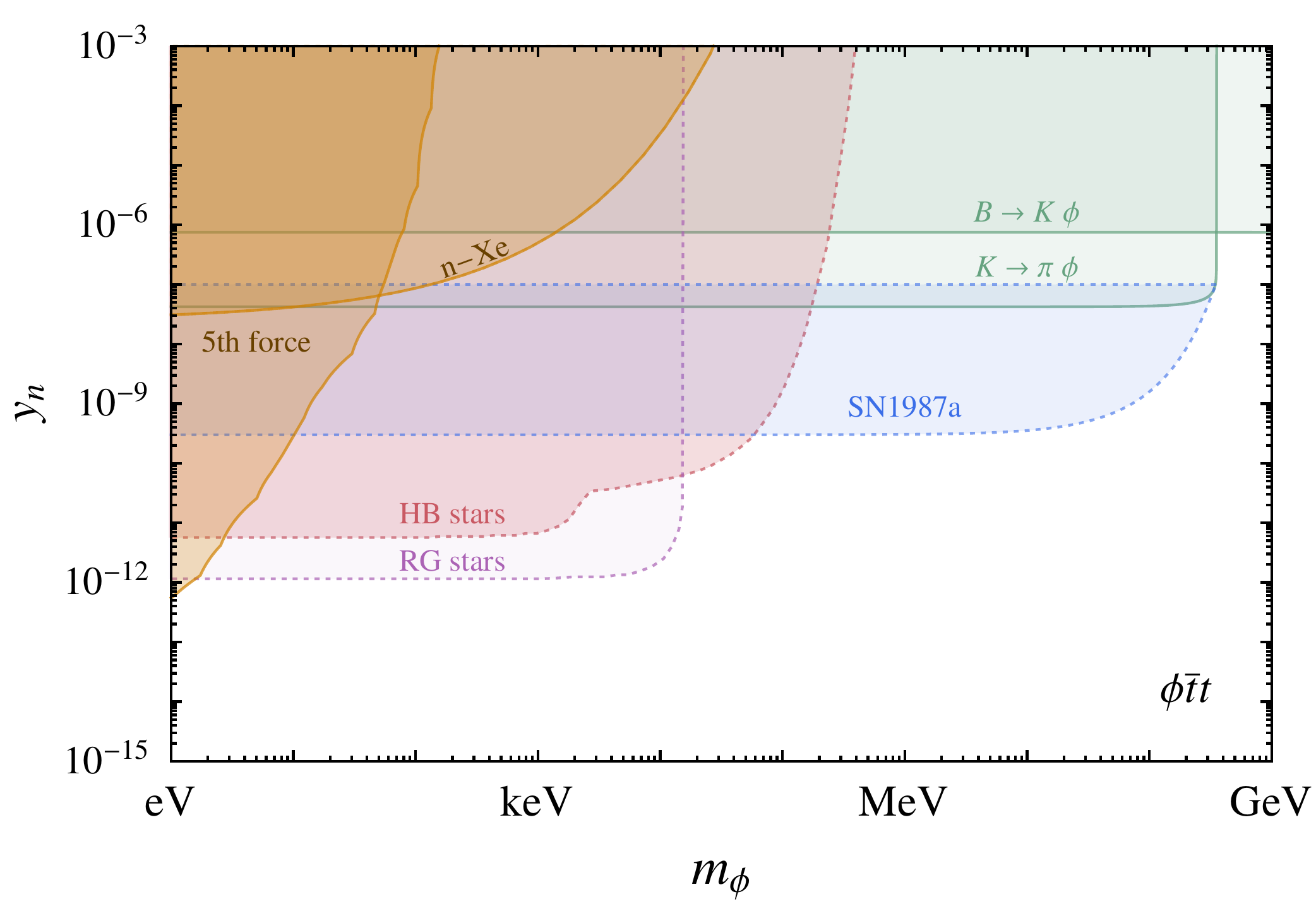} \\
\includegraphics[width=0.75\textwidth]{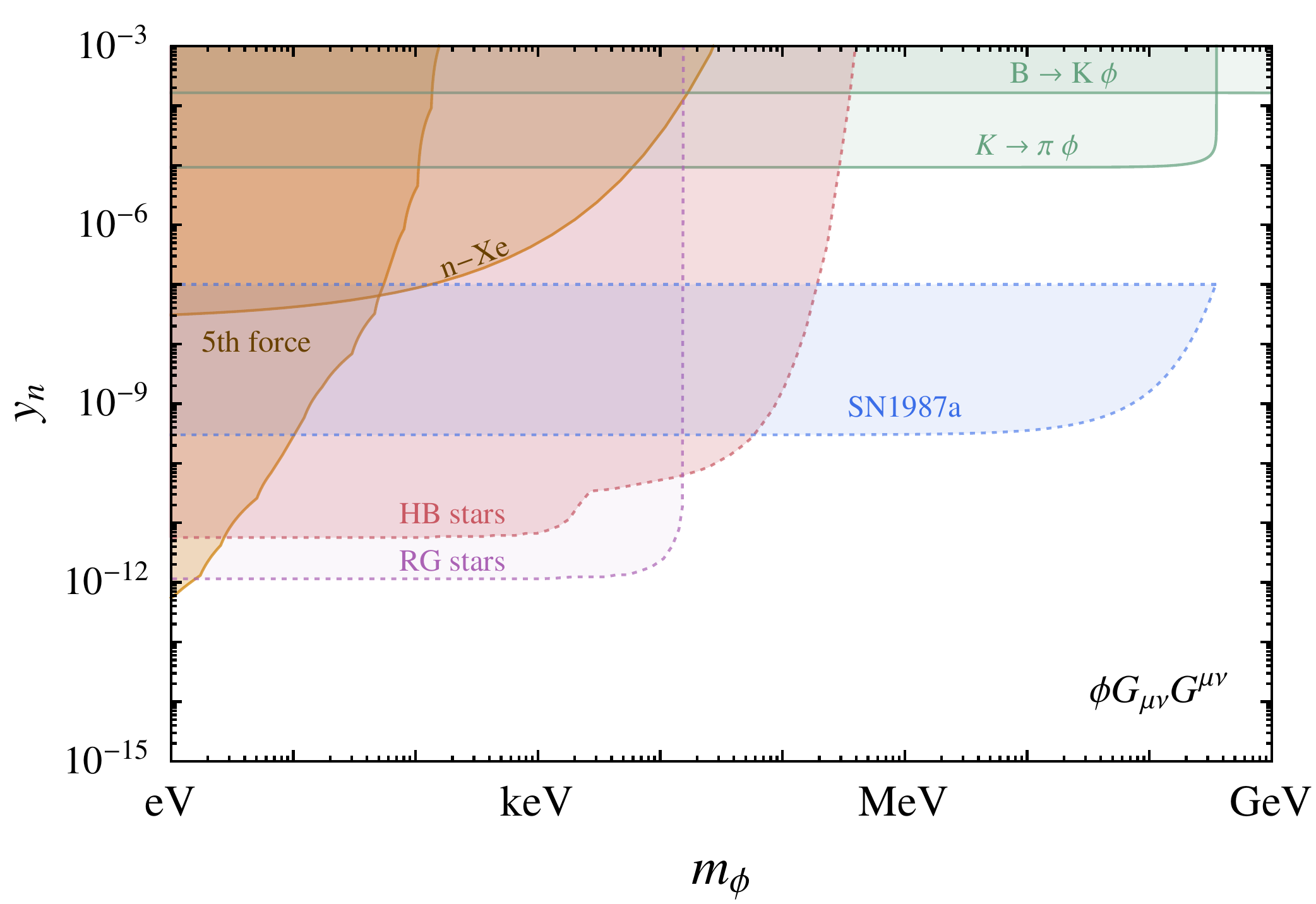}
\caption{ Constraints on a sub-GeV scalar mediator, given in terms of the effective scalar-nucleon coupling $y_n$. In the top panel, we assume that the nucleon interaction is generated by a $\phi$-top coupling and in the bottom panel, we assume it is generated by a $\phi$-gluon coupling (for instance from a heavy colored fermion). We show limits from fifth force \cite{Murata:2014nra} and neutron scattering searches \cite{Leeb:1992qf} (orange), rare meson decays (green),  and stellar cooling limits from HB stars \cite{Hardy:2016kme} (red), RG stars \cite{Hardy:2016kme} (purple) and SN1987A (blue). \label{fig:mphiplane} }
\end{figure}

In this work, we consider two possible origins of the nucleon interaction in Eq.~\eqref{eq:nucleonmodel}: one with the mediator coupling to the top, and one with the mediator coupling to a vectorlike generation of heavy, colored particles. This distinction is only important for bounds from meson decays, which are primarily sensitive to the coupling of $\phi$ with top quarks.
\begin{enumerate}
\item First consider a coupling to top quarks of the form
\begin{equation}\label{eq:topcoupling}
\mathcal{L}\supset\frac{\epsilon}{\sqrt{2}} \frac{m_t}{v} \phi t\bar t
\end{equation}
with $v=246$ GeV. This in turn induces the gluon coupling
\begin{align}\label{eq:gluoncoupling}
\mathcal{L}\supset\frac{\alpha_s}{4\Lambda}\phi G^a_{\mu\nu}G^{a\mu\nu}\qquad\mathrm{with}\quad \frac{1}{\Lambda}=\frac{\epsilon}{3\pi v},
\end{align} 
which at low energies maps to a nucleon coupling
 \begin{equation}\label{eq:mesonmaptop}
\mathcal{L}\supset y_n\phi \bar n n \qquad\mathrm{with}\quad y_n=-\epsilon\frac{2 m_n}{3b v}\approx - 2.6\times 10^{-4}\epsilon,
\end{equation}
where $b=29/3$ is the first coefficient of the QCD $\beta$-function. Here we assumed that $m_\phi$ is below the strange quark mass, and neglected the small light quark contributions to the nucleon mass~\cite{Gunion:1989we}. While we do not explicitly consider couplings to the lighter quarks, they would not substantially change the constraints, provided that $\phi$ does not mediate a large flavor-changing interaction. For example, a model where $\phi$ couples to the SM quarks through minimal flavor violation (MFV) would be consistent with our setup, though we here assume that the coupling to leptons can be neglected.
 
 \item Alternatively, if the mediator couples to a colored vector-like generation, we also generate the gluon interaction in Eq.~\eqref{eq:gluoncoupling}, where $\Lambda$ is now a function of the mass ($m_Q$) and coupling of the heavy generation. We assume for simplicity that these additional particles are outside the reach of the LHC, with $m_Q\approx 5$ TeV and $\Lambda$ a free parameter. The map to the low energy theory is given by
 \begin{equation}
	 y_n=-\frac{2\pi}{ b}\frac{m_n}{\Lambda}\approx- 0.65\frac{m_n}{\Lambda}.
	\label{eq:vectorlikequarks}
 \end{equation}

Although there is no direct coupling to the top quark at tree-level, Eq.~\eqref{eq:topcoupling} is still generated radiatively and can lead to rare meson decays. The leading contribution is 
 \begin{equation}
 \epsilon  \approx \sqrt{2}\frac{\alpha_s^2}{\pi}\log \left(\frac{m_Q^2}{m_t^2}\right)      \frac{v}{\Lambda}.   
 \end{equation}
Similar couplings to the lighter quarks are generated as well, but are less relevant to the meson constraints we consider here. 
This induced coupling to the top quark can also be written as $\epsilon \approx -17 y_n$.
 \end{enumerate}
As we will see, the meson constraints depend on $\epsilon$ rather than $y_n$. For a fixed value of $y_n$, they are therefore weaker in the model with a vector-like generation. 
The compiled constraints on $\phi$ for these two scenarios are summarized in Fig.~\ref{fig:mphiplane}, and are described in detail below.

\subsection{Terrestrial constraints \label{sec:nucleon_constraints}}

\subsubsection{Meson constraints\label{sec:mesons}}

\begin{figure}[t]
\includegraphics[width=0.4\textwidth]{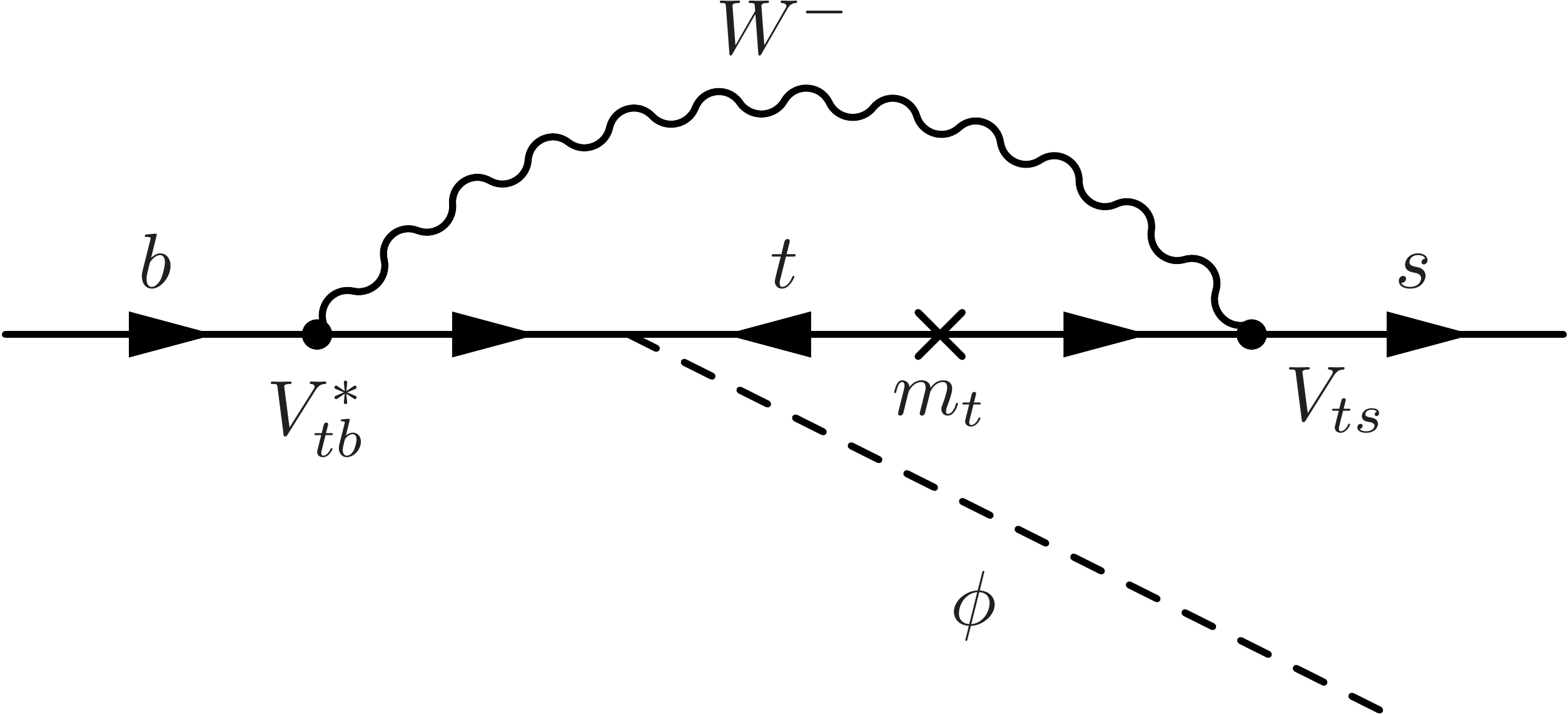}\hfill
\includegraphics[width=0.4\textwidth]{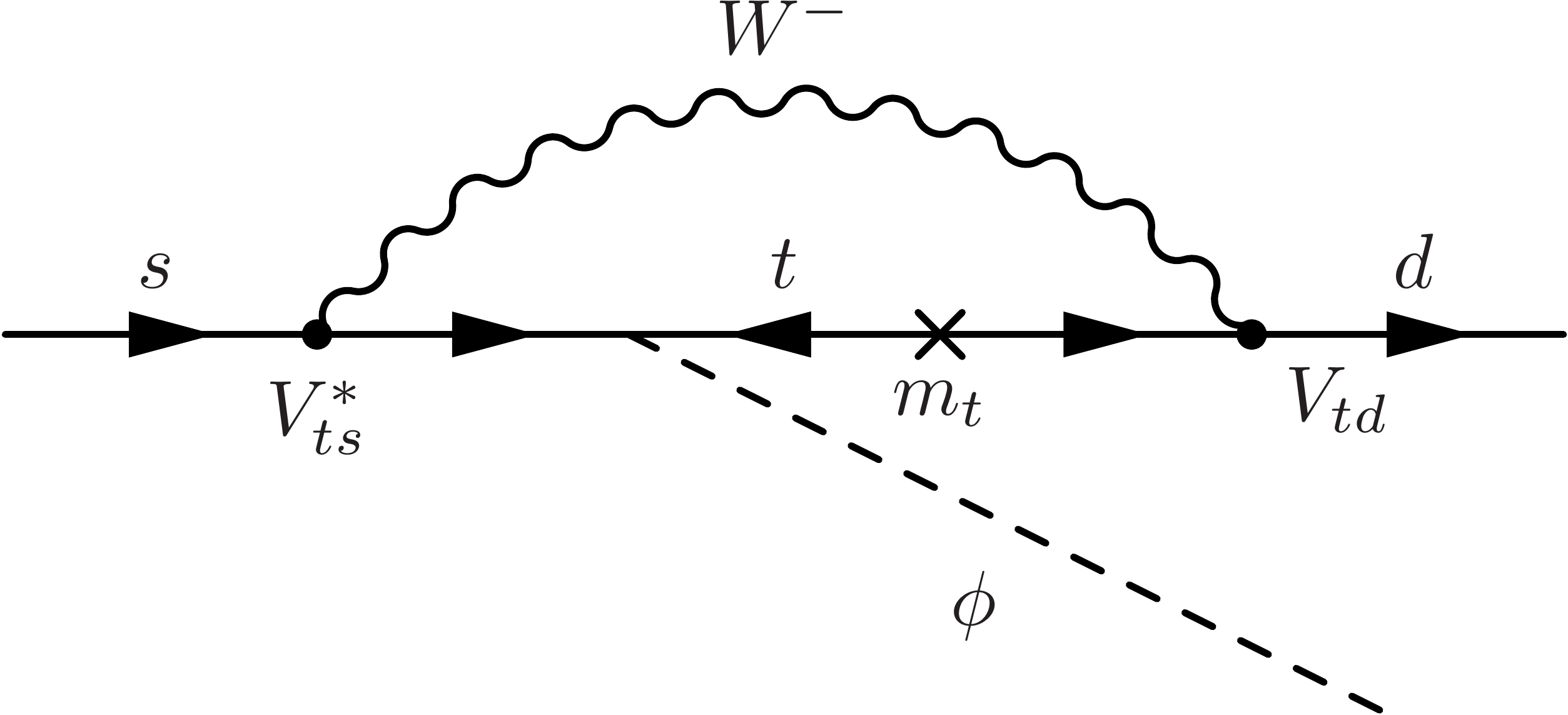}
\caption{Diagrams generating $B\rightarrow K \phi$ (left) and $K\rightarrow \pi \phi$ (right). \label{fig:mesondecay} }
\end{figure}

A light scalar with a coupling to hadrons appears in exotic meson decays, like $B\rightarrow K \phi$ and $K\rightarrow \pi \phi$. Generally, the dominant contribution to these decay rates comes from a top-$W$ loop (see Figure \ref{fig:mesondecay}), which explains the need to specify the origin of the nucleon coupling in Eq.~\eqref{eq:nucleonmodel}.

In both models outlined above, the (indirect) coupling to the top quark opens up the radiative decay $\phi\rightarrow \gamma\gamma$, with width of
\begin{equation}\label{eq:photonwidth}
\Gamma_{\phi\rightarrow\gamma\gamma}=\frac{q_t^4 N_c^2\alpha^2}{144\pi^3}\frac{m_\phi^3}{v^2}\epsilon^2
\end{equation}
with $q_t=2/3$ and $N_c=3$. In the absence of a competing decay mode, this results in a lifetime for $\phi$ of
\begin{equation}
c\tau \approx \left(\frac{\mathrm{MeV}}{m_\phi}\right)^3\times \frac{1}{\epsilon^2}\times 10^{8}\, \mathrm{cm}.
\end{equation}
We will therefore always treat $\phi$ as missing energy in these decays, regardless of whether it decays to photons or to an invisible state ({\em e.g.} the DM).  The relevant flavor measurements to compare with are:
\begin{equation}
\begin{array}{lll}
Br(B\rightarrow K\, \nu\bar\nu  )& <1.6\times 10^{-5} &\text{\cite{Lees:2013kla}}\\
Br(K\rightarrow \pi\, \nu\bar\nu ) &=1.73^{+1.15}_{-1.05}\times10^{-10}&\text{\cite{Artamonov:2008qb}}.
\end{array}
\end{equation}
The partial width for $B\rightarrow K \phi$ is given by (e.g.~\cite{Willey:1982mc}),
\begin{equation}
\begin{array}{ll}
\Gamma_{B\rightarrow K \phi}&=\frac{|C_{sb}|^2 f_0(m_\phi)^2}{16\pi m_{B}^3}\left(\frac{m_B^2-m_K^2}{m_b-m_s}\right)^2\xi(m_B,m_K,m_\phi)\\
C_{sb}&=\frac{3 m_b m_t^2 V^\ast_{ts}V_{tb}}{16\pi^2 v^3}\epsilon \quad\quad \xi(a,b,c)=\sqrt{(a^2-b^2-c^2)^2-4b^2c^2}\\
f_0(q)&=0.33(1-q^2/38 \mathrm{GeV}^2).
\end{array}
\end{equation}
where $f_0(q)$ parametrizes the hadronic form factor \cite{Ball:2004ye}.

The width for $K \rightarrow \pi \phi$ follows the same expression, with the appropriate substitution of the masses and CKM matrix elements. The form factor for this process is well approximated by $f_0(q)\approx 1$ in the low $q$ limit \cite{PhysRevD.53.R1}.  For each scenario, the resulting bounds on $\epsilon$ can then be converted to a bound on $y_n$. For $m_\phi \ll m_\pi$, the strongest bounds come from the $K\rightarrow \pi \phi$ process:
\begin{align}
y_n \lesssim 4.2\times 10^{-8}& \qquad\qquad\qquad \text{(top coupling)}\\
y_n \lesssim 9.3\times10^{-6}& \qquad\qquad\qquad \text{(vector-like generation)}.
\end{align}
The constraints are shown in green on Figure \ref{fig:mphiplane}.

\subsubsection{$5^{\mathrm{th}}$ force constraints\label{sec:5thforce_nucleon}}

For mediator masses below $\sim$ 100 eV, various $5^{\mathrm{th}}$ force experiments become relevant (see {\em e.g.}~\cite{Murata:2014nra,Adelberger:2003zx} for recent reviews.). This is important in particular for the cosmological fate of $\phi$: at least for values of $y_n$ that can give rise to detectable values of the direct detection cross section, $5^{\mathrm{th}}$ force constraints prevent $\phi$ from being light enough to behave as dark radiation at late times \cite{Green:2017ybv,Hochberg:2015pha}. We will comment more on the cosmology in Section~\ref{sec:cosmo}.

At low energies, the presence of the massive, scalar mediator introduces an attractive Yukawa potential between two macroscopic objects, of the form
\begin{equation}
V(r)=-\frac{y_n^2}{4\pi}\frac{1}{r}e^{-m_\phi r}
\end{equation}
per nucleon pair.
$5^{\mathrm{th}}$ force constraints are conventionally parametrized in terms of a modification to the gravitational potential
\begin{equation}
V(r)=-G_N\frac{m_1 m_2}{r}\left(1+\alpha e^{-m_\phi r}\right)
\end{equation}
with $m_{1,2}$ the test masses in the potential, $G_N$ Newton's constant and $\alpha$ parametrizing the strength of the additional force. Since $m_n\gg m_e$, and since $\phi$ couples to all nucleons, we make the identification
\begin{equation}
\alpha =\frac{y_n^2}{4\pi}\frac{M_{\textrm{pl}}^2}{m_n^2}.
\end{equation}
In particular, for the mediator masses most relevant for us, the strongest constraints arise from Casimir force experiments~\cite{Murata:2014nra}, and are shown as the orange region in Figure \ref{fig:mphiplane}.  For $m_\phi\approx 1$ eV and  $y_n \lesssim  10^{-12}$ the bound crosses over to the stellar constraints discussed below.

In addition, bounds on new forces with masses as heavy as an MeV can also be obtained from low-energy neutron experiments~\cite{Barbieri:1975xy,Nesvizhevsky:2007by,Kamiya:2015eva}.  We show limits from neutron-Xe scattering, derived in Ref.~\cite{Kamiya:2015eva}, giving a limit $y_n < 10^{-7}$ for $m_\phi \ll $ MeV. The dark matter itself can mediate a force through virtual effects, which implies that there is still a (weaker) constraint for $m_\chi\ll m_\phi$, even if the mediator itself is too short ranged \cite{Fichet:2017bng}.

\subsection{Astrophysical constraints\label{sec:astro_nucleon}}

\subsubsection{Stellar emission\label{sec:stellar_nucleon}}

Light bosons with small couplings to electrons or nucleons can be emitted in stars, giving rise to rapid cooling. New energy loss processes are constrained in a number of stellar systems, giving strong limits on the coupling of the light boson (see \cite{Raffelt:1996wa} for a review). Here we consider limits from horizontal branch (HB) stars, red giants (RG) and supernova 1987A (SN1987A).

Horizontal branch (HB) and red giant (RG) stars have temperatures close to $T \approx 10$ keV, required for helium burning in the core. Bosons of mass up to $\sim 10-100$ keV can thus be emitted in the core, escaping the star and leading to a new form of energy loss. The lifetime of horizontal branch stars is measured by the ratio of the abundance of red giant to that of horizontal branch stars, and would be shortened depending on the energy loss rate $\epsilon$. Existing constraints on bosons with small couplings to nucleons primarily utilize a condition $\epsilon \lesssim 10$ erg/g/s. This approximate condition applies both for HB and RG stars; in the latter case, the constraint is due to the fact that additional energy loss can delay the onset of helium ignition. For a detailed discussion, see Ref.~\cite{Raffelt:1996wa}. 

More massive mediators can be constrained by the luminosity of SN1987A, where the core temperature was around $T \approx 30$ MeV. Here the requirement is that any new energy loss satisfies $\epsilon \lesssim 10^{19}$ erg/g/s~\cite{Raffelt:1996wa}.
In addition, due to the large core density  ($\rho \sim 10^{15}$ g/cm$^3$ rather than $\rho \sim 10^{4}$ g/cm$^3$ in HB stars), light bosons emitted in the core may be re-absorbed before escaping. This leads to a trapping regime, where the coupling of the bosons is large enough that they do not efficiently escape the core. In this regime, the new particles can still modify energy transport within the star and may be constrained, but this requires detailed modeling beyond the scope of this work.

{\bf Bounds from SN1987A.} 
For a scalar with coupling $y_n \phi \bar n n$, constraints from SN1987A were derived in the weak-coupling limit in Ref.~\cite{Ishizuka:1989ts}. Following these results, we require that the energy loss per unit mass be $\epsilon < 10^{19}$ erg/g/s; taking a fiducial set of parameters $T = 30$ MeV and $\rho = 3 \times 10^{14}$ g/cm$^3$, this gives a limit of $y_n \lesssim 10^{-10}$. (For $m_\phi$ close to $T$, we simply assume a Boltzmann suppression of $e^{-m_\phi/T}$ in the rate.) This bound does not apply to large $y_n$ due to the trapping effect discussed above -- light scalars can be re-absorbed on nuclei with a mean free path smaller than the core. If $\phi$ decays to dark matter, then the decay length may be much shorter than the $\phi$ re-absorption mean free path. Then the question of whether the energy is lost depends on the mean free path of the dark matter. As we will see in the next section, for $m_\phi>2m_\chi$ scenario we must typically require that $\chi$ does not thermalize with $\phi$ in the early universe to evade BBN bounds. This puts a sufficiently stringent upper bound on $y_\chi$ such that the $\phi\rightarrow \chi\chi$ decay is not relevant for SN1987A.

Ref.~\cite{Ishizuka:1989ts} did not provide a calculation of trapping via scalar re-absorption. We estimate that trapping is relevant for $y_n \gtrsim 10^{-7}$ simply by taking the results for axions~\cite{Turner:1987by,Ishizuka:1989ts,Burrows:1990pk}. Our justification is the following: despite the different parametrics of scalar and axion production, the weak-coupling constraints on axions and on scalars are quite similar, with $y_n \lesssim 5 \times 10^{-11}$ for axions. Since the production rate and absorption rates are related by detailed balance, $\Gamma_{\textrm{prod}}(\omega) = e^{-\omega/T} \Gamma_{\textrm{abs}}(\omega)$, we find to leading order that the absorption mean free path is given by $\ell_{\textrm{abs}}^{-1} \propto \epsilon \, \rho /T^4$ with $\epsilon$ the energy loss rate~\cite{Raffelt:1996wa}.   Hence, we expect the ratio of the $y_n$ at the trapping boundary to $y_n$ at the weak-coupling limit to be similar for both axions and scalars.

There are a number of caveats in the bounds above, aside from the estimate of the trapping regime we have used. First, the weak-coupling result in Ref.~\cite{Ishizuka:1989ts} was obtained with a simplified model of the SN core, and the result can vary by up to an order of magnitude depending on the core temperature and radius. For instance, see Ref.~\cite{Chang:2016ntp} for a discussion of systematic uncertainties for SN1987A bounds on dark photons. In addition, the dominant production mode  in this case is nucleon-nucleon scattering with $\phi$ emission. Existing results have been calculated with the approximation of one-pion exchange. As discussed in subsequent work~\cite{Hanhart:2000ae,Hanhart:2000er,Rrapaj:2015wgs}, one-pion exchange is not an accurate description of nucleon-nucleon scattering data; instead, they used a soft theorem description along with nucleon-nucleon scattering data to calculate production rates, leading to differences of up to an order of magnitude in some models.  Finally, the result of Ref.~\cite{Ishizuka:1989ts} did not account for production due to mixing with the longitudinal component of the photon, an effect discussed in Ref.~\cite{Hardy:2016kme}. 

{\bf Bounds from HB and RG stars.} 
Constraints on scalars coupling to baryons from stellar emission were given in Refs.~\cite{Grifols:1986fc,Grifols:1988fv}, with $y_n \lesssim 4.3 \times 10^{-11}$ from HB stars~\cite{Raffelt:1996wa}. This result was derived assuming the Compton process $\gamma + \textrm{He} \to \textrm{He} + \phi$ for $m_\phi \lesssim 10$ keV. Recently, constraints on light scalars were updated in Ref.~\cite{Hardy:2016kme}, which included the effects of in-medium mixing on the production of scalars. To summarize, the scalar can mix with longitudinal photon polarization  modes in a star, leading to an additional contribution to the rate. This production mechanism is possible as long as $m_\phi < \omega_p$, where the plasma frequency $\omega_p$ is also the oscillation frequency of the longitudinal mode. For horizontal branch stars $\omega_p \sim 2$ keV and for red giants $\omega_p \sim 20$ keV. We use the constraints given in Ref.~\cite{Hardy:2016kme}, which come from production in HB stars via brem off ions (``continuum'') and mixing effects (``resonant''), and from production in RG cores via mixing effects.  It is also possible that for sufficiently large couplings the scalars may be trapped in RG and HB stars as in the case of SN1987A, though this requires detailed modeling of the energy transport in the star~\cite{Raffelt:1988rx,Carlson:1988jg}.  For RG and HB stars, the couplings at which trapping would likely be relevant are also in a regime where the terrestrial constraints become important, so we do not consider this possibility further.

\subsubsection{Dark matter self interactions \label{sec:selfint}}

When $\chi$ composes all of the dark matter, there are significant constraints on dark matter self-interactions from the shapes of halos (see Ref.~\cite{Tulin:2017ara} for a review).  Because of the low momentum transfer involved in the scattering, the self-interaction constraints on the dark matter coupling, $\alpha_\chi=y_\chi^2/4\pi$, are particularly strong in the limit of a light mediator with $m_\phi \lesssim m_\chi v$ and $v \sim 10^{-3}$.
Bullet-cluster and halo shape observations tell us that DM self-interactions should satisfy
\begin{equation}
	\frac{\sigma}{m_\chi} \lesssim 1-10 \mbox{ cm}^2/\mbox{g},
\end{equation}
where $\sigma$ is the self-interaction cross section. 

For scattering of distinguishable particles, the relevant cross section is the transfer cross section $\sigma_T$, which is the scattering weighted by momentum transfer (see Appendix~\ref{app:SIDM}). However, for the particular model at hand with identical particles, we instead use the viscosity cross section, defined as
\begin{equation}
	\sigma_V = \int d\Omega \frac{d \sigma}{d \Omega}\sin^2\theta,
\end{equation}
in order to regulate the forward and backward scattering divergences~\cite{Tulin:2013teo}. 
For our benchmark model, the non-relativistic Born cross section is 
\begin{equation}
\sigma_V^{\text{born}}\approx\frac{\alpha_\chi^2 \pi}{m_\chi^2 v^4}\left(\frac{R^4+2R^2+2}{R^2(R^2+2)}\log\left[1+R^2\right]-1\right)
\end{equation}
with $v$ the relative velocity and $R=m_\chi v / m_\phi$.
In the heavy mediator limit with $R \ll1 $ and cross section bound of $1 \mbox{ cm}^2/\mbox{g}$, the corresponding constraint on the coupling constant is
\begin{align}
	\alpha_\chi \lesssim  0.025 \left(\frac{1 \mbox{ keV}}{m_\chi}\right)^{1/2} \left(\frac{m_\phi}{1 \mbox{ MeV}}\right)^2.
\end{align}
Meanwhile, for light mediators with $R\gg1$,
\begin{align}
	\sigma_V^{\text{born}} \approx \frac{\alpha_\chi^2 \pi}{m_\chi^2 v^4}\left(\log R^2 -1\right).
\end{align}
For instance, taking $v = 10^{-3}$ and $R=10$, we have
\begin{align}
	\alpha_\chi \lesssim 6\times 10^{-10} \times  \left(\frac{m_\chi}{1 \mbox{ MeV}}\right)^{3/2}.
\end{align}

Furthermore, assuming dark matter self-interactions are responsible for the deviations of observed halo shapes from $\Lambda$CDM, it is possible to {\em fit} the interaction cross section to the shapes of dwarf galaxies, elliptical galaxies, and clusters; this was carried out in Ref.~\cite{Kaplinghat:2015aga}.  Remarkably, they found that a mediator mass on the 1-10 MeV scale (depending on the dark matter mass) was favored by the data.

 The self-interaction constraints are substantially relaxed when $\chi$ is only a fraction of the dark matter. In addition, the effects of a strong self-interaction may enter a new regime where the dark matter behaves as a fluid~\cite{Ahn:2002vx,Ahn:2004xt,Agrawal:2016quu}. This would form an additional isothermal component of the Milky Way's dark matter halo. For the parameter space we consider with light (but not massless) mediators, we expect that dissipative effects are kinematically suppressed by finite $m_\phi$.  Thus, as a representative case, we will take $\Omega_\chi/\Omega_{\textrm{DM}} \approx 0.05$ in relaxing the SIDM constraints~\cite{Cyr-Racine:2013fsa}. Interestingly, partially interacting dark matter may also have some connections with some discrepancies in large scale structure measurements~\cite{Chacko:2016kgg,Buen-Abad:2017gxg}.

\subsection{Cosmology \label{sec:cosmo}}

\newcommand{\Tphi}{T_{\phi,\mathrm{dec}}}
\newcommand{\Tchi}{T_{\chi,\mathrm{th}}}

The detailed thermal history of the universe must be addressed in scenarios where the dark matter and/or mediator are relativistic during BBN and recombination, since the light particles may contribute to the effective number of relativistic degrees of freedom $\Neff$. In the standard model, $\Neff \approx 3.046$. The deviation from the standard model value can be written as
\begin{equation}
	\Delta \Neff=\frac{4}{7} \sum_i g_i \left(\frac{T_i}{T_\nu}\right)^4
\end{equation} 
with $g_i$ and $T_i$ the effective degrees of freedom and the temperature of the various relativistic species in the dark sector, and $T_\nu$ the neutrino temperature after electron decoupling in the standard cosmology.  

In general, the strongest bounds come from CMB constraints on light degrees of freedom present at late times, but they can vary significantly depending on the assumed cosmology and data sets included in the fit. The most stringent constraint from CMB (namely {\emph{Planck}}) and large scale structure gives $ \Neff^{\textrm{CMB}} = 3.04 \pm 0.18 (1 \sigma)$~\cite{Ade:2015xua}. However, this fit is for the minimal extension of the standard cosmology, and is modified in the presence of other physics. For instance, if an additional eV-scale sterile neutrino is included, the 95$\%$ CL constraints weaken to $\Neff^{\textrm{CMB}} < 3.7, \ \ m_{\nu, {\rm sterile}}^{{\rm eff}} < 0.38\, \textrm{eV}$, while if the mass of active neutrinos is included, the 95$\%$ CL constraint from CMB plus large scale structure is $\Neff^{\textrm{CMB}} =3.2 \pm 0.5, \ \ \sum m_{\nu} < 0.32\, \textrm{eV}$. 
Another modification to the standard picture is if neutrinos have a somewhat large self interaction -- instead of free-streaming, they may behave as a fluid at late times. For instance, Ref.~\cite{Oldengott:2017fhy}  found that $\Neff^{\textrm{CMB}} = 3.0 \pm 0.3 (1 \sigma)$ in this scenario which again increases the uncertainty. Finally, as noted in Ref.~\cite{Cyburt:2015mya}, the bounds presented by the {\emph{Planck}} collaboration generally assume a particular relationship between the He fraction and the baryon asymmetry.

In this paper, we choose to be as agnostic as possible about the detailed cosmological history, such that the most robust bounds on our model come from BBN only. In particular, a recent combined fit of He and D abundances from Ref.~\cite{Cyburt:2015mya}, driven by improved errors in the measured D abundance, gives $\Neff^{\textrm{BBN}} = 2.89 \pm 0.28 (1 \sigma)$. Using the 2D likelihood for $\Neff$ and the baryon-to-photon ratio $\eta$ shown in  Ref.~\cite{Cyburt:2015mya}, the 2$\sigma$ bound on $\Neff$ is
\begin{equation}\label{eq:BBNbound}
\Delta N_{\rm eff}^{\rm BBN}\lesssim 0.5,
\end{equation}
which implies roughly 2$\sigma$ tension with a single real scalar with a temperature similar to that of the neutrinos ($\Delta N_{eff}^{BBN}\approx 0.57$). 
In similar spirit, Ref.~\cite{Cyburt:2015mya} finds that the CMB constraint is
\begin{equation}
\Delta N_{\rm eff}^{\rm CMB}\lesssim 0.6
\end{equation}
 at recombination, where this bound uses only CMB data and no assumptions about the He fraction from BBN are made.
 The next stage of CMB experiments can significantly improve on these results, with a projected sensitivity of $\sigma( \Neff^{\textrm{CMB}} ) \approx 0.04 $ from CMB Stage IV~\cite{Abazajian:2016yjj}.

Both low mass dark matter and mediators may contribute to $\Delta \Neff$. 
In this section, we are considering the most optimistic case where both the dark matter and the mediator are real scalars, such that $g_\phi=g_\chi=1$. If the dark sector was ever in thermal contact with the SM, it is also necessary to introduce a mechanism to avoid $\phi$ and/or $\chi$ having too large an abundance:
for $m_\phi \gtrsim 10$ eV, the relic abundance of $\phi$ becomes a non-negligible component of the dark matter and for much larger masses it exceeds the observed density of DM. (For $m_\phi \approx 1$ eV, close to the boundary of the fifth force constraints, $\phi$ behaves as hot DM, but is less than 1$\%$ of the dark matter.) 

In the case where $\phi$ becomes thermalized with the SM, the simplest solution for its relic abundance is to introduce an additional light degree of freedom, as also considered in \cite{Green:2017ybv}. This additional degree of freedom can also be instrumental in setting the relic abundance of $\chi$.\footnote{There are a number of other ways one could reduce any excess density of non-relativistic $\phi$ and/or $\chi$. First, due to its (indirect) coupling with the top quark, $\phi$ has a radiative decay to photons, given in Eq.~\eqref{eq:photonwidth}.  While this process can lead to the decay of $\phi$ before it becomes non-relativistic, one needs $y_n\gtrsim 10^{-7}\times \sqrt{1\,\MeV/m_\phi}$ \mbox{($y_n\gtrsim 4 \times10^{-6}\times \sqrt{1\,\MeV/m_\phi}$)} for the $\phi \bar t t$ ($\phi GG$) coupling model. This is only satisfied in a very small part of the parameter space allowed by the meson constraints. It is also {\em a priori} possible to have $\phi$ decay to neutrinos via a $\phi \bar\nu \nu$ portal. This is viable for $m_\phi\gtrsim 10\,$ MeV, and excluded for $m_\phi\lesssim 10$ MeV if $\phi$ is in equilibrium with the neutrinos during BBN~\cite{Boehm:2013jpa}. For $m_\phi \ll 1\, \MeV$ it is also possible that the mediator enters equilibrium only after BBN but before $\phi$ becomes non-relativistic. 
} We therefore extend our simplified model with a real scalar $a$ with couplings of the form
\begin{equation}
	\mathcal{L}\supset -\frac{1}{2} m_a  a^2-\frac{1}{2}y_a m_a \phi a^2-\frac{1}{4}\lambda \chi^2 a^2.
	\label{eq:lightdof}
\end{equation}
We take $m_a \ll \, $eV such that $a$ is a small contribution to the energy density at late times, with the main constraints coming from CMB $\Neff$ bounds. 

In order to determine $\Neff$, we must determine the ratio of the temperature of the dark sector relative to the neutrino temperature. This quantity depends on the cosmological history, specifically on the temperature at which the dark sector decoupled from the standard model bath, as well as the number of dark degrees of freedom that were in equilibrium when this decoupling occurred. In particular, for sufficiently small $y_a$, it is possible that $a$ does not come into equilibrium until after the dark sector decouples from the SM. At that point, $a$ can then be responsible for the relic abundance of $\phi$ and $\chi$. For instance, as long as $y_a  \lesssim 10^{-1} ( \eV/m_a)$, then $a$ is not in equilibrium with $\phi$ until after $\phi$ decouples from the SM bath (which occurs at $T \approx m_\pi$, as we discuss below). On the other hand, the decay $\phi \to a a$ is in equilibrium through the $\phi$ mass threshold for
\begin{align}
		y_a \gtrsim 10^{-5} \left( \frac{m_\phi}{ 200\, \keV} \right)^{3/2} \left( \frac{\eV}{m_a} \right),
\end{align}
allowing for efficient depletion of the $\phi$ abundance. 
Therefore, in this scenario our estimates of $\Delta \Neff^{\rm BBN}$ depend only on whether $\phi$ and $\chi$ have equilibrated with the SM. 

\begin{figure}[t]
\centering
\includegraphics[width=0.49\textwidth]{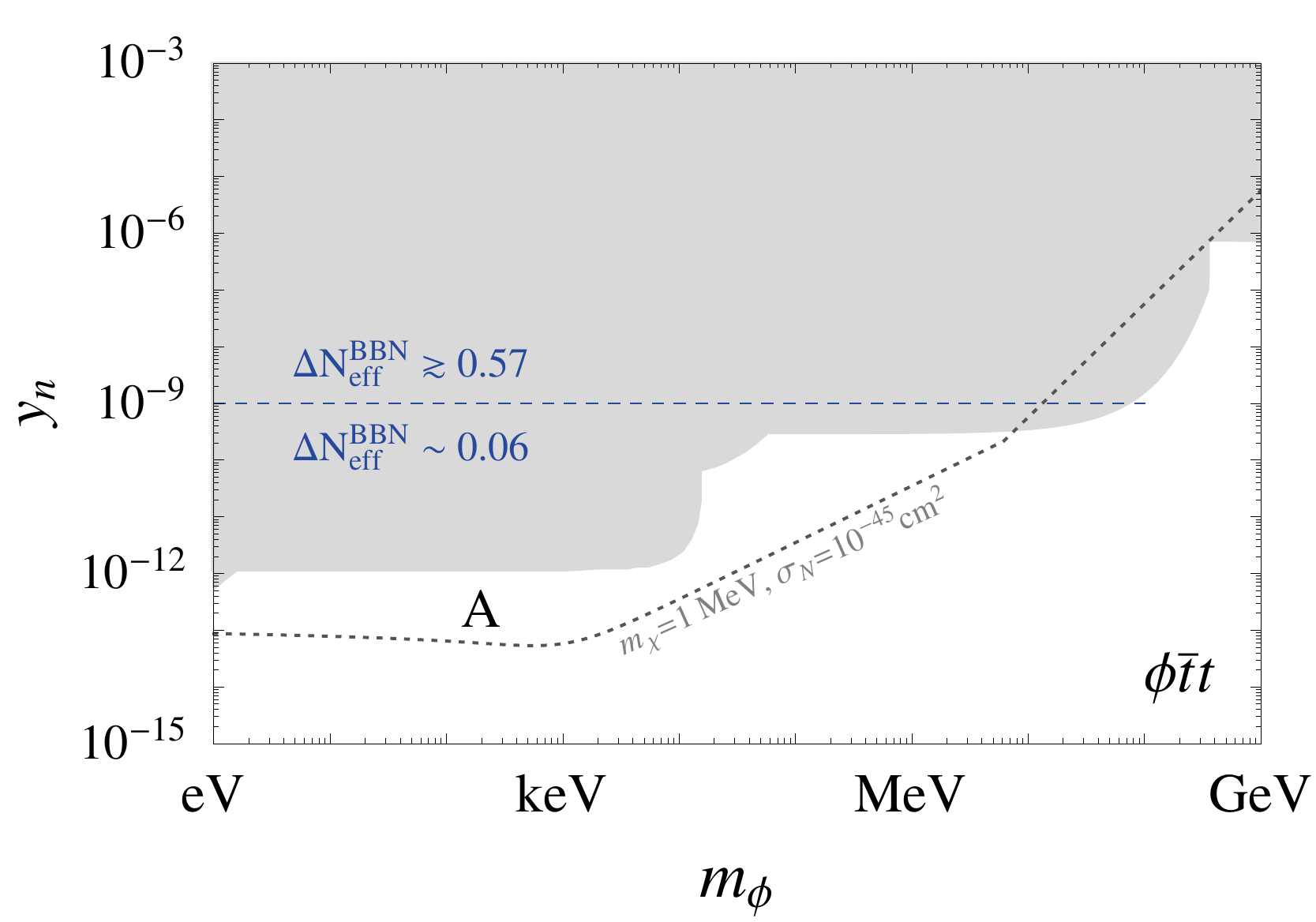}\hfill
\includegraphics[width=0.49\textwidth]{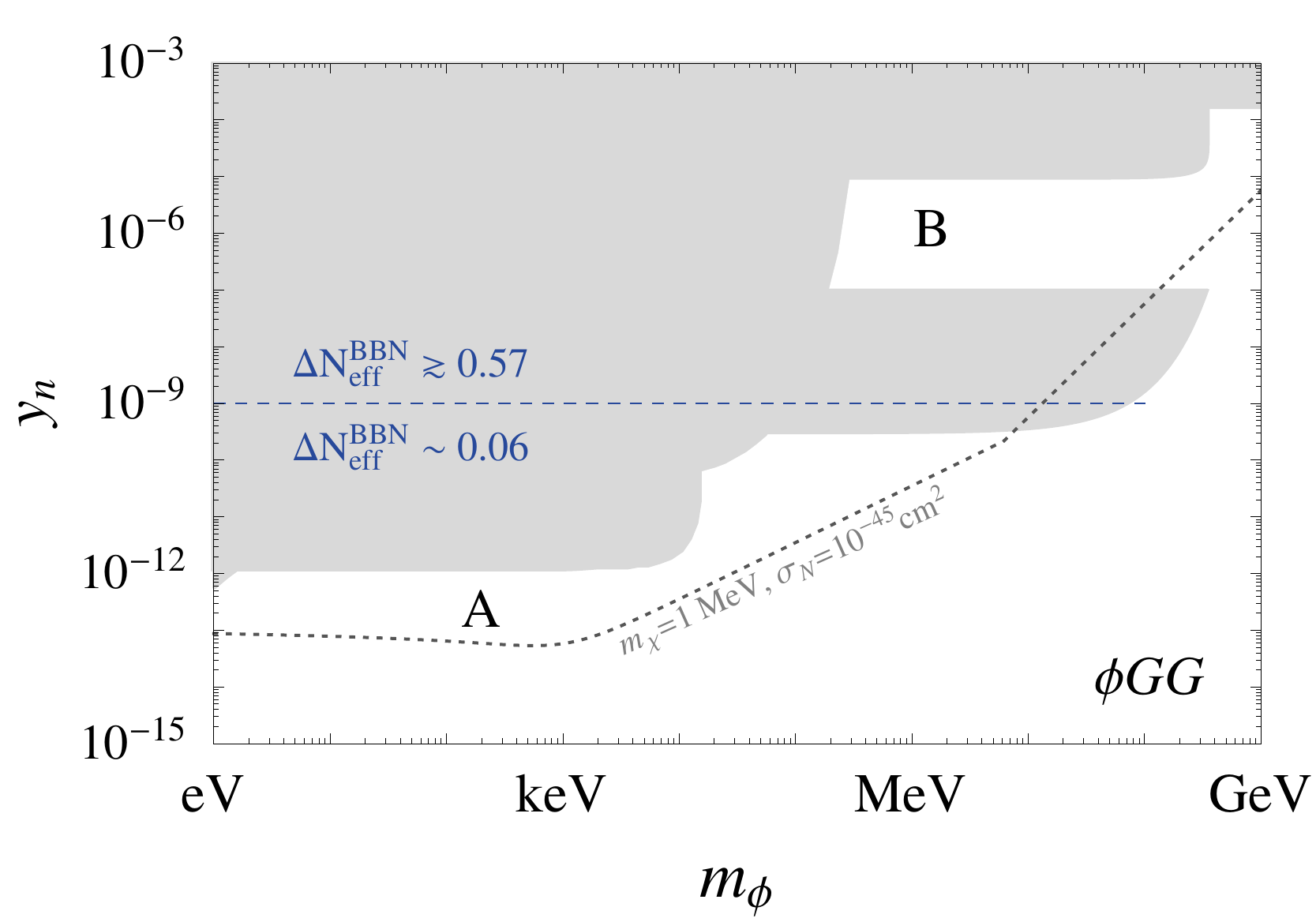}
\caption{ Thermalization history in relation to the $m_\phi$ vs  $y_n$ plane, where the gray shaded regions are the constraints from Fig.~\ref{fig:mphiplane}. Below the dashed blue line the dark sector decouples from the SM before the QCD phase transition. Approximate values of $\Delta N_{\rm eff}$ are shown for both regions (see text). A representative cross section contour for fixed $m_\chi$, and where $\alpha_\chi$ is chosen to saturate self-interaction bounds, is indicated by the dotted gray line.
\label{fig:thermalizationoverview} }
\end{figure}

Above the QCD phase transition $ T_{\rm QCD} \approx 300$ MeV, from Eqs.~(\ref{eq:gluoncoupling}-\ref{eq:mesonmaptop}) we can write the coupling of $\phi$ with gluons in terms of $y_n$,
\begin{align}\label{eq:gluoncoupling2}
	\frac{\alpha_s b y_n}{8 \pi m_n} \phi G_{\mu \nu}^a G^a_{\mu \nu},
\end{align}
where we assume the $\phi$ coupling to light quarks is negligible. Thermal scatterings such as $g g \to \phi g$ can bring the mediator into equilibrium. Since the coupling in \eqref{eq:gluoncoupling2} is given by an irrelevant operator, the mediator drops out of equilibrium with the SM as the universe cools. Estimating the cross section as $\sigma \propto \frac{\alpha_s^3 b y^2_n}{64 \pi^2 m_n^2} $, we find this process is out of equilibrium by $T = 300$ MeV for $y_n \lesssim 10^{-9}$. (For more detailed estimates of these rates, see Appendix~\ref{app:therm}.) This qualitative boundary is shown by the dashed blue line in Fig.~\ref{fig:thermalizationoverview}, in relation to the terrestrial and astrophysical constraints discussed in the previous sections. In the supernova trapping window, we see that the mediator always remains in equilibrium (region B). This is only relevant for the $\phi GG$ model, where the meson constraints are somewhat weaker. The dashed gray line is intended to give some intuition on the possible direct detection cross sections, which will be discussed in more detail in the next section. 

Then, taking $y_n \ll 10^{-9}$ (region A in Fig.~\ref{fig:thermalizationoverview}) such that the mediator decoupled before the QCD phase transition, the contribution to $\Neff$ from the dark sector is at most
\begin{align}
	\Delta \Neff^{\textrm{BBN}} \approx  \frac{4}{7} \sum_i g_i \left( \frac{g_{SM}(T_{\nu dec})}{g_{SM}(T_{QCD})} \right)^{4/3} \approx   0.06 \ \sum_i g_i
	\label{eq:Neff_A}
\end{align}
where we took $g_{SM}(T_{\rm QCD}) \approx 61.75$ and $g_i$ the degrees of freedom of the species in the dark sector which are in equilibrium before decoupling.  This case is unconstrained with current CMB or BBN data but may be probed by CMB Stage IV \cite{Abazajian:2016yjj}, depending on the value of $g_{SM}$ at the temperature where the dark sector decouples. For values of $y_n$ consistent with the stellar constraints, the dark sector decouples from the SM at $T\approx 100$ TeV, such that there is plenty of room for a suitable mechanism to set the dark sector relic density. In particular, if $\chi$ is assumed to be all of the dark matter, then annihilation of $\chi \chi \to \phi \phi$ is not sufficient to obtain the correct relic abundance -- this is because of the bounds on $y_\chi$ from dark matter self-interactions. However, the annihilation $\chi \chi \to a a$ can set the correct abundance if
\begin{align}
	\lambda \approx 3 \times 10^{-7} \left( \frac{m_\chi}{\MeV} \right).
	\label{eq:relicdensity_lambda}
\end{align}
Self-interaction bounds can still be satisfied, since $\chi-\chi$ scattering is only generated through a loop of $a$ particles and thus is higher order in $\lambda$, scaling as $\lambda^4$ while the annihilation cross section scales as $\lambda^2$.  Since this way of setting the relic density via thermal freeze-out in the dark sector is essentially independent of the direct detection and other phenomenology, we choose to be agnostic about the specific mechanism whenever possible.

If $y_n \gtrsim 10^{-9}$ (region B in Fig.~\ref{fig:thermalizationoverview}), $\phi$ remains in equilibrium throughout the QCD phase transition and even afterwards due its coupling with pions/nucleons. While the scattering on nucleons is suppressed by the baryon-to-photon ratio $\eta \approx 6\times 10^{-10}$, the mediator still scatters with pions. In particular, the gluon coupling in \eqref{eq:gluoncoupling2} induces a pion coupling~\cite{Gunion:1989we},
\begin{align}
	\frac{b y_n}{18 m_n} \phi \, \partial \pi^\dagger \partial \pi.
\end{align}
For $y_n \gtrsim 10^{-9}$, this keeps $\phi$ in equilibrium until around $T \approx m_\pi$ when pions decouple.
In this case, since pion and muon decoupling occurs more or less simultaneously, it is conservative to assume that the dark sector has the same temperature as the neutrinos. This implies
\begin{align}
	\Delta \Neff^{\textrm{BBN}} \approx  \frac{4}{7} \sum_i g_i,
	\label{eq:Neff_B}
\end{align}
summing over the dark degrees of freedom $g_i$ just below pion decoupling. If $\chi$ and $\phi$ are both in equilibrium at this time, $\Delta \Neff^{\textrm{BBN}}\approx 1.14$, which is firmly excluded. If only $\phi$ is in thermal contact with the SM we have $\Delta \Neff^{\textrm{BBN}}\approx 0.57$ and $\Delta \Neff^{\textrm{CMB}}\approx 0.72$,  both of which are in roughly 2$\sigma$ tension with current data. The $\Delta \Neff^{\textrm{CMB}}$ number was obtained by transferring the energy density of $\phi$ to the light scalar $a$, where we assumed that $a$ and $\phi$ were not in equilibrium with one another until after $T \approx m_\pi$.
 In what follows, we will consider the 2$\sigma$ tension associated with $\phi$-SM equilibrium as permissible, but we will insist that the dark matter $\chi$ does not thermalize with the mediator.  Its relic density must therefore have a different origin, such as from the interaction with the scalar $a$ (as discussed above), or from interactions with the neutrinos \cite{Berlin:2017ftj}.  

\begin{figure}[t]
\centering
\includegraphics[width=0.6\textwidth]{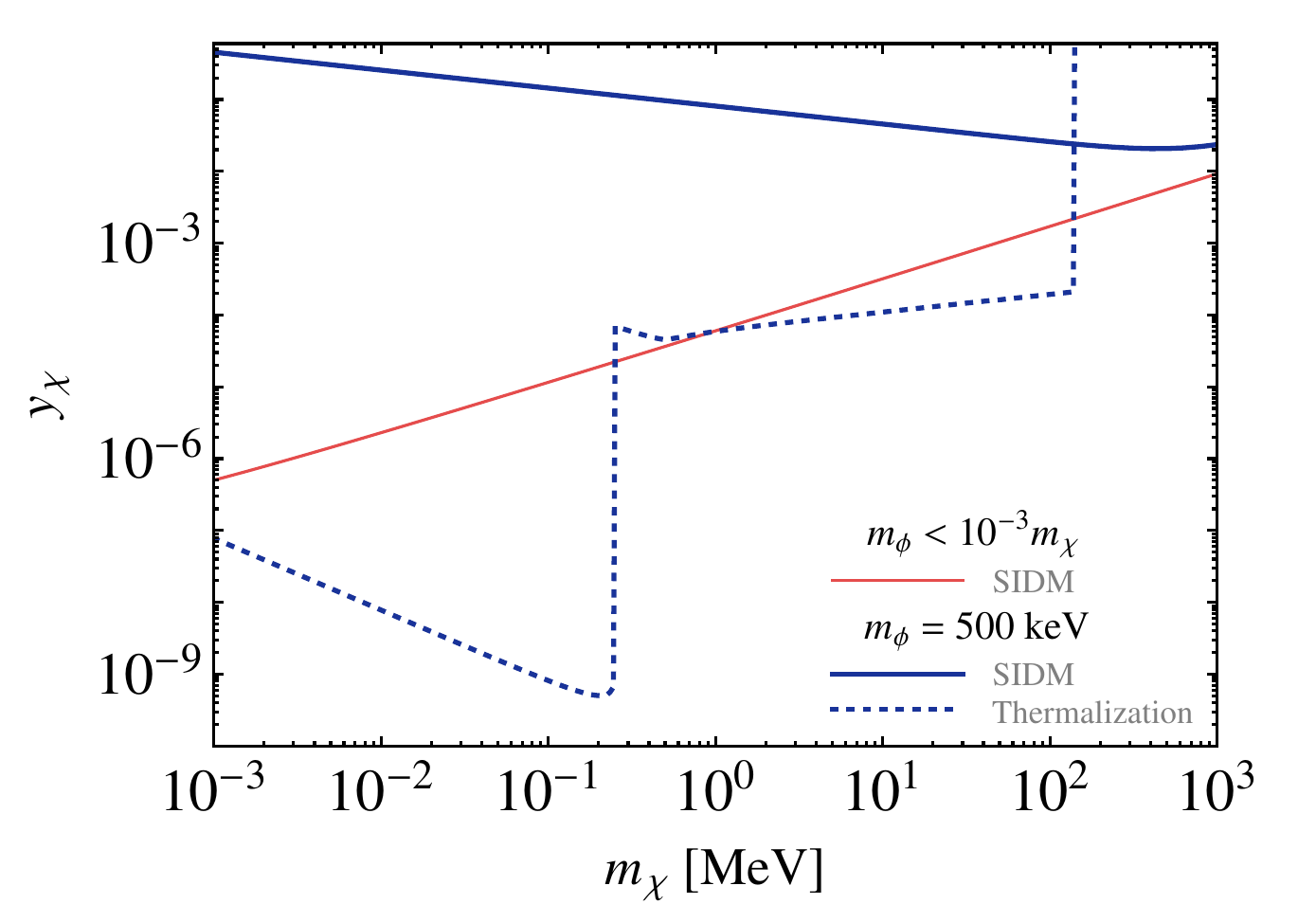}
\caption{ In the $y_\chi$ vs $m_\chi$ plane, we compare limits from self-interactions (assuming $\Omega_\chi/\Omega_{\textrm{DM}} = 1$) to our bound on thermalization of $\chi$ with $\phi$, using the conditions given in Eq.~\eqref{eq:phichieqb}. In the light mediator case ($m_\phi < 10^{-3} m_\chi$), we do not place a bound on $\chi$ thermalization since the values of $\Neff$ satisfy current bounds even with 2 degrees of freedom. For the massive mediator benchmark ($m_\phi = 500$ keV), and for couplings above $y_n \approx 10^{-9}$,  we require that at most $\phi$ is in equilibrium with the SM to avoid $\Neff$ bounds; this gives the thermalization bound shown (dashed blue line). Above $m_\chi = T_\pi$, we assume that the dark matter can annihilate away efficiently and deposit entropy back into the pions.
\label{fig:alphaXtherm}
}
\end{figure}

Before turning to the direct detection prospects, we first discuss the implications of forbidding $\phi$-$\chi$ equilibrium. Possible $\chi$ thermalization mechanisms are annihilation of $\phi \phi \to \chi \chi$ or $\phi \rightarrow \chi\chi$ decay, with the latter possible only if $m_{\phi}>2m_\chi$. If the decay is open, it dominates the thermalization process, since it is lower order in $y_\chi$. The thermalization conditions at a particular temperature $T$ are
\begin{equation}
\begin{array}{ll}
\frac{m_\phi}{T}\Gamma_{\phi\rightarrow\chi\chi} \sim H(T)&\quad\quad \text{if }m_\phi>2m_\chi\\[0.75em]
n_\phi(T) \langle v\sigma_{\phi\phi\rightarrow\chi\chi}\rangle \sim H(T)&\quad\quad \text{if }m_\phi<2m_\chi,
\end{array}
\label{eq:phichieqb}
\end{equation} 
where the factor $m_\phi/T$ in the decay rate accounts for the Lorentz boost of $\phi$ at high temperatures. We evaluate these conditions at $T\approx \text{Max}[m_\phi,m_\chi]$ or $T\approx m_\pi$, whichever is lower. As long as $\phi$ is relativistic, we have $n_\phi(T)\approx 0.38\, T^3$. In all expressions, we neglected thermal corrections to the potential, which is justified as long as $y_\chi^2 T\ll m_{\phi}$. The decay rate is given by
\begin{equation}
	\Gamma_{\phi \rightarrow \chi\chi}=\frac{y_\chi^2}{32\pi}\frac{m_\chi^2}{m_\phi} \sqrt{1-4m_\chi^2/m_\phi^2}.
\end{equation}
and, in the limit where $m_{\chi,\phi}\ll\sqrt{s}$, the cross section is 
\begin{equation}
	\sigma_{\phi\phi\rightarrow\chi\chi}\approx \frac{y_\chi^4}{16\pi}\frac{ m_\chi^2}{s^2}.
\end{equation}
With $s \sim T^2$, it is clear that both decay and scattering become more important compared to Hubble as the temperature drops, so that $\chi$ will enter equilibrium as the universe cools. For our numerical results, we use the full expression for $\sigma_{\phi\phi\rightarrow\chi\chi}$ (without expanding in $m_{\chi,\phi}\ll\sqrt{s}$) and numerically evaluate the thermally averaged cross section \cite{Gondolo:1990dk}
\begin{equation}\label{eq:thermaver}
	\langle \sigma_{\phi \phi \to \chi \chi} v \rangle=\frac{1}{8m_\phi^4 T \big(K_2(m_\phi/T)\big)^2}\int^{\infty}_{4m^2_\phi}\!\!ds\, \sigma \,(s-4m_\phi^2)\sqrt{s} K_1(\sqrt{s}/T),
\end{equation}
where $K_{1,2}$ are modified Bessel functions of the second kind. The resulting constraint on $y_\chi$, as derived from Eq.~\eqref{eq:phichieqb}, as well as the self-interaction constraint of Section~\ref{sec:selfint}, are shown in Fig.~\ref{fig:alphaXtherm} for two benchmark points. The feature around $m_\phi = 2m_\chi$ clearly indicates where $\phi\rightarrow\chi\chi$ decay becomes relevant. We do not impose a $\phi$-$\chi$ thermalization constraint for $m_\chi\gtrsim m_\pi$, since in this case the energy density of $\chi$ can still be deposited into the SM sector rather than in dark radiation. As can be seen in Fig.~\ref{fig:alphaXtherm}, for a massive mediator in the supernova trapping window (region B in Fig.~\ref{fig:thermalizationoverview}), the thermalization constraint is almost always dominant over the self-interaction constraint. Instead, for the light mediator limit (region A in Fig.~\ref{fig:thermalizationoverview}), self-interaction bounds are important.

Finally, we note that $\Delta \Neff$ may be additionally suppressed in more elaborate models, as has been studied in some detail for light sterile neutrinos. Possible examples are late time entropy production \cite{Gelmini:2004ah,Chacko:2016hvu}, non-conventional cosmological evolution of the mass parameters \cite{Zhao:2017wmo,Ghalsasi:2016pcj} or a late dark sector phase transition \cite{Chacko:2004cz}. This may remove the need for demanding that the DM does not equilibrate, and could open more parameter space.

\subsection{Results \label{sec:results_neutron}}

As suggested by the results in Figs.~\ref{fig:mphiplane} and \ref{fig:thermalizationoverview}, terrestrial and astrophysical constraints indicate two possible regimes where direct detection of sub-MeV dark matter is conceivable in this simplified model:
\begin{itemize}
\item  $m_\phi\ll m_\chi$  and $y_n\lesssim 10^{-12}$ (Region A in Fig.~\ref{fig:thermalizationoverview}): The dark sector decouples from the SM before the QCD phase transition, which cools the dark sector relative to the SM sector enough such that $\Delta \Neff$ satisfies current bounds (Eq.~\eqref{eq:Neff_A}). SIDM constraints, shown in  Fig.~\ref{fig:alphaXtherm}, provide the strongest constraints on $y_\chi$ assuming that $\chi$ is the dominant component of the dark matter.

\item $m_\phi \gtrsim 500 \,\keV$ and $10^{-5}\gtrsim y_n\gtrsim 10^{-7}$ in the $\phi GG$ model (Region B in Fig.~\ref{fig:thermalizationoverview}): The mediator $\phi$ is in equilibrium with the SM until the pion threshold, leading to $\Delta \Neff \approx \tfrac{4}{7}$. This is roughly in 2$\sigma$ tension with BBN and CMB constraints. To avoid firm exclusion from BBN/CMB, we further require that $\chi$ cannot thermalize with $\phi$, which restricts $y_\chi$ as shown in Fig.~\ref{fig:alphaXtherm}.
\end{itemize}

\begin{figure}[p]
\includegraphics[width=0.49\textwidth]{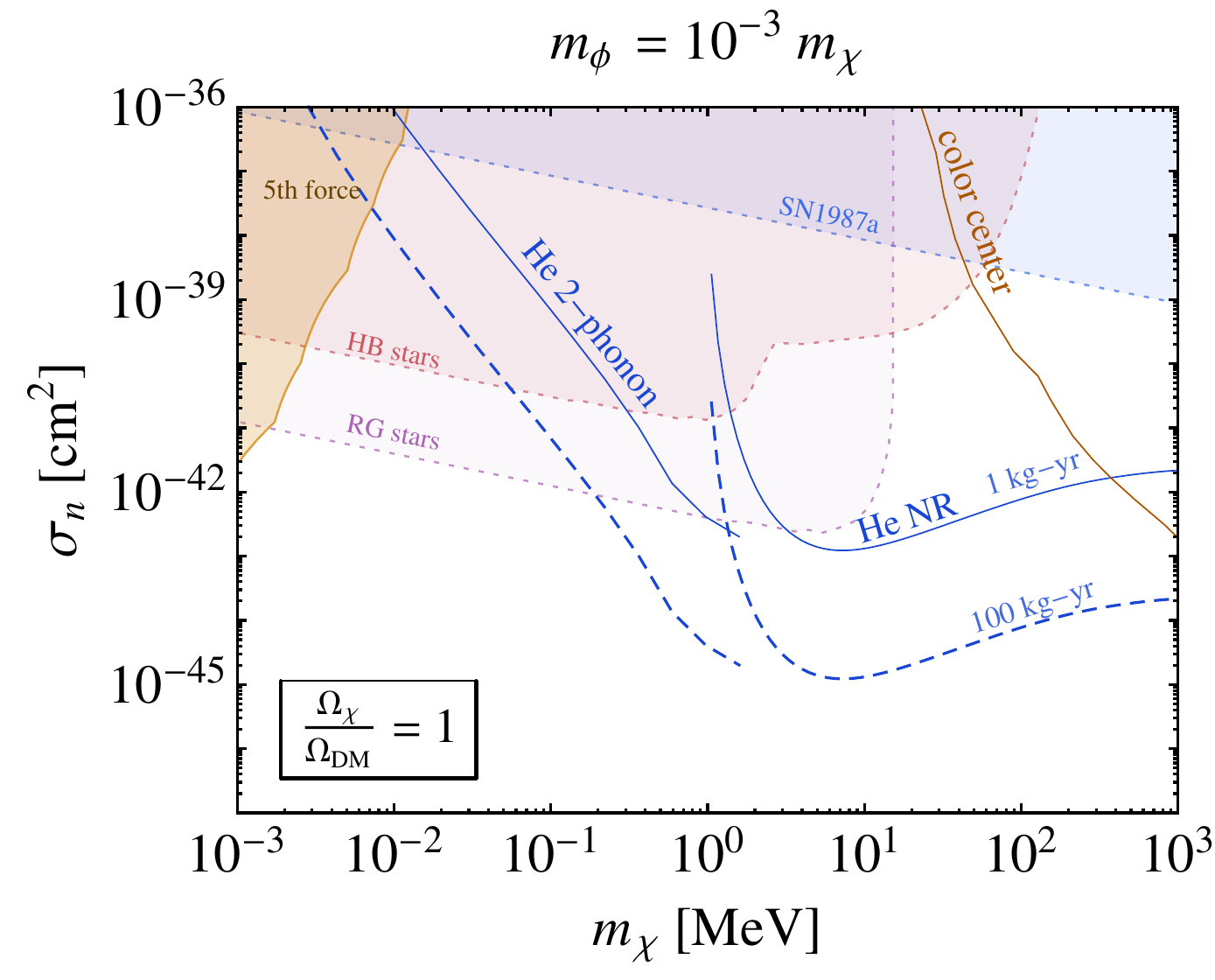}\hfill
\includegraphics[width=0.49\textwidth]{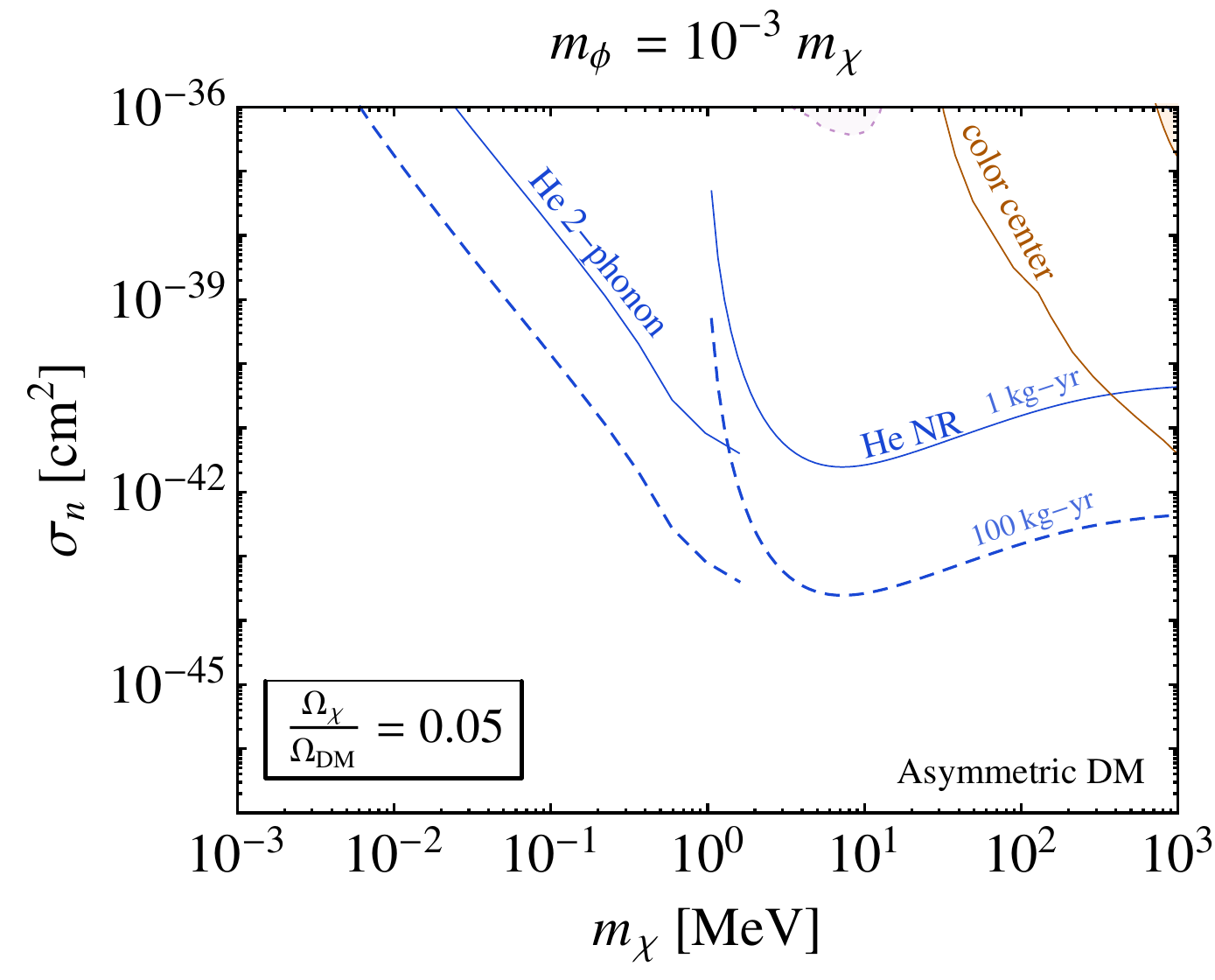}
\caption{  
Direct detection cross section as function of the dark matter mass, where the mediator mass is $m_\phi=10^{-3} m_\chi$. We show the case that $\chi$ is all the dark matter (left) and where $\chi$ composes $5\%$ of the dark matter (right). In the former case $y_\chi$ is fixed by saturating the self-interaction constraint, while in the latter case we take $y_\chi = 1$ and assume $\chi$ is a complex scalar with an asymmetric relic abundance. The blue lines indicate the projected reach with superfluid helium in the multi-phonon and nuclear recoil modes~\cite{Knapen:2016cue}, assuming that the nuclear recoil mode includes energies from 3 meV up to 100 eV. We also show projected reach for color centers~\cite{Budnik:2017sbu}, where in this case we show their sensitivity for the massless mediator limit. \label{fig:scenA_money}
}
\end{figure}

\begin{figure}[p]
\includegraphics[width=0.49\textwidth]{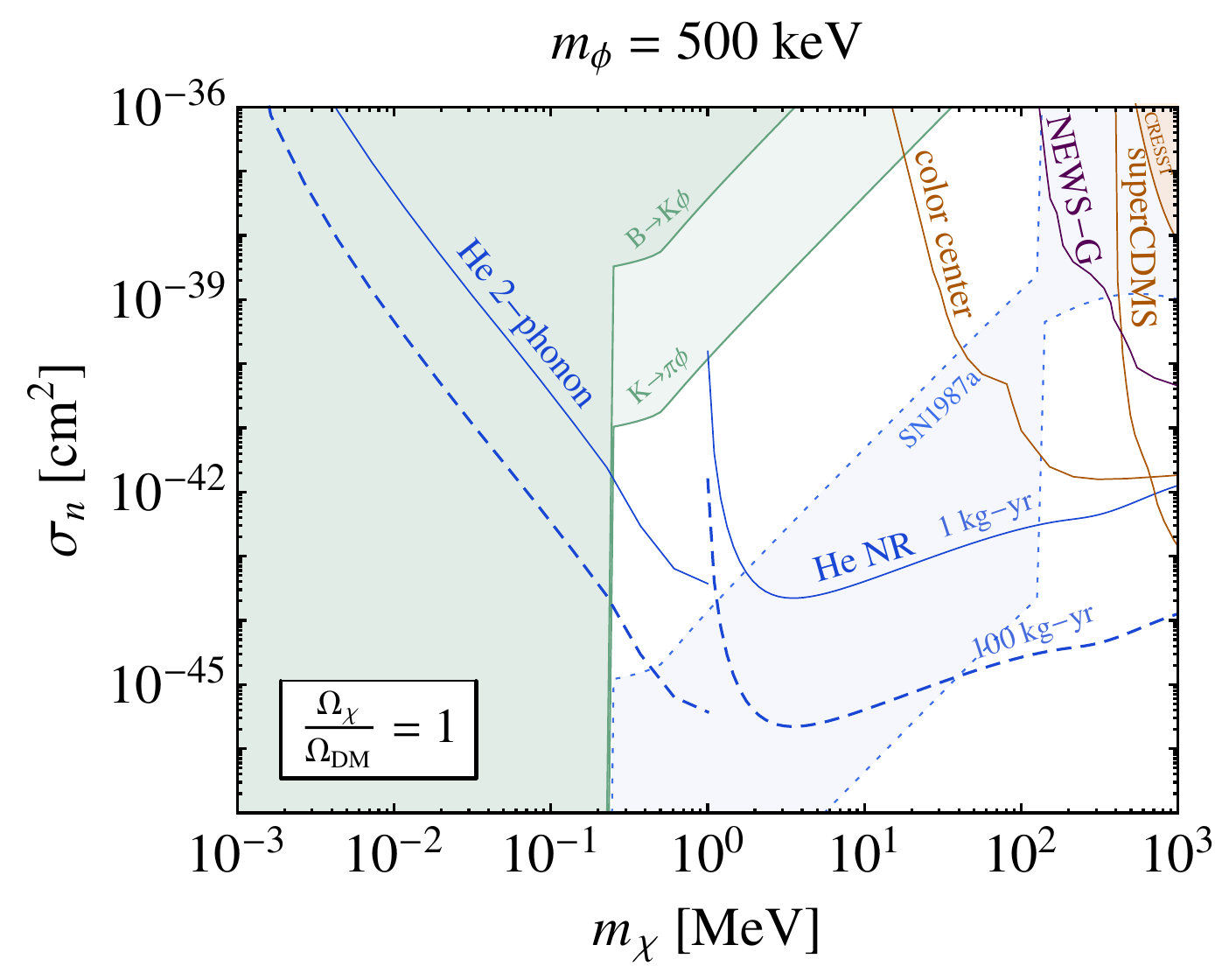}\hfill
\includegraphics[width=0.49\textwidth]{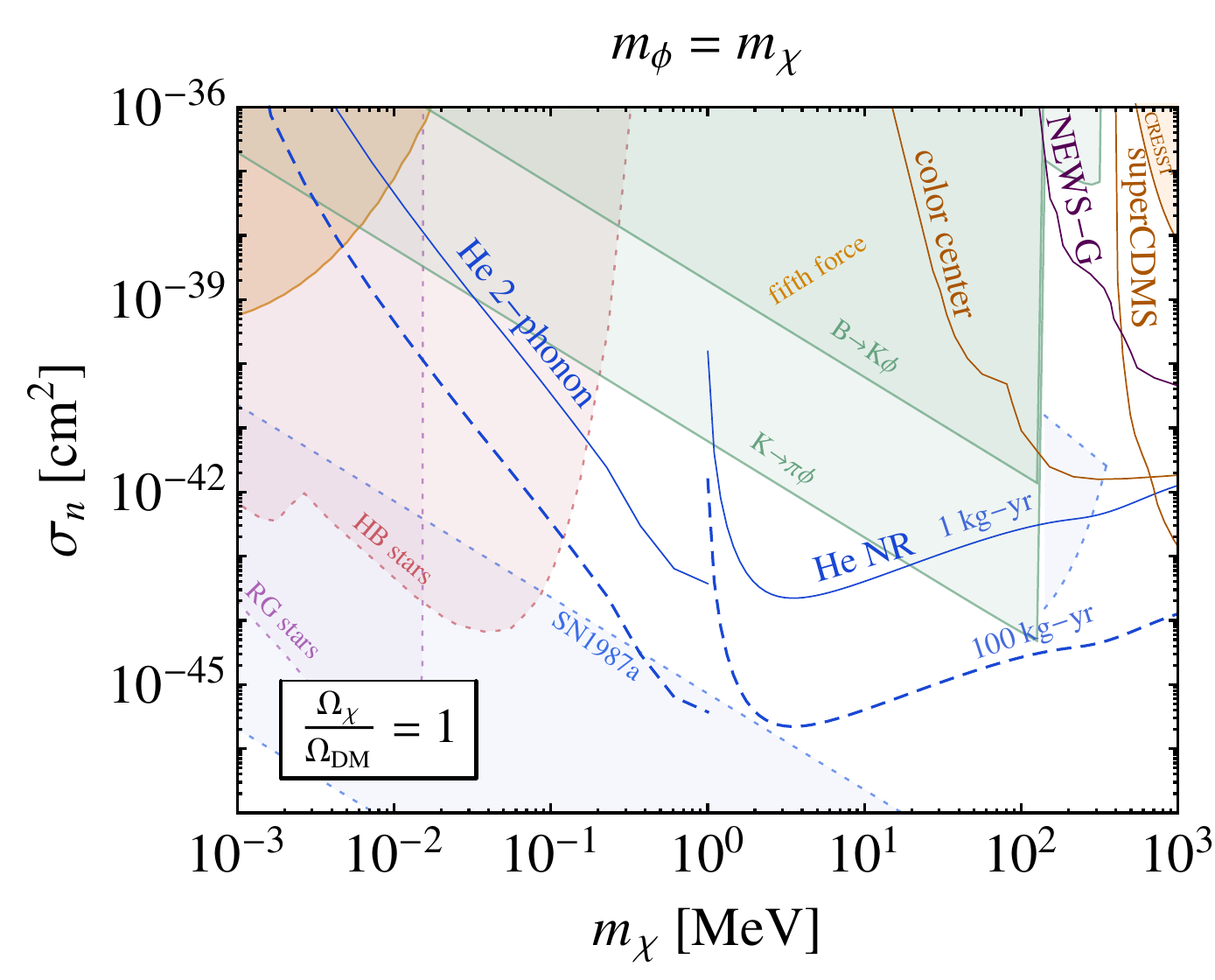}
\caption{  Constraints and detection prospects in the direct detection cross section vs $m_\chi$ plane, for the heavy mediator regime. We scan over $y_n$ and fix $y_\chi$ by demanding that the dark matter does not thermalize with $\phi$ for $m_\chi<m_\pi$. For $m_\chi > m_\pi$, the self-interaction constraint is used instead. We show the projected reach for NEWS-G and SuperCDMS~\cite{Battaglieri:2017aum}, as well as proposed experiments with superfluid helium~\cite{Knapen:2016cue} or color centers~\cite{Budnik:2017sbu}.  The orange shaded region is excluded by CRESST~\cite{Angloher:2015ewa}. For all of the accessible direct detection for $m_\chi < 100$ MeV, we note that $\phi$ is in equilibrium with the SM until after the QCD phase transition and thus $\Delta \Neff \approx \tfrac{4}{7}$, which is in $\approx 2\sigma$ tension with current BBN and CMB bounds. \label{fig:scenB_money}}
\end{figure}

In Fig.~\ref{fig:scenA_money}, we present direct detection prospects for $m_\phi \ll m_\chi$, fixing the $m_\phi/m_\chi$ ratio. The existing constraints are compared with the reach for superfluid helium~\cite{Knapen:2016cue} and color centers~\cite{Budnik:2017sbu}. We saturate self-interaction constraints and include stellar bounds and fifth force constraints. If $\chi$ composes all of the dark matter in the light mediator regime, direct detection with near future experiments appears to be challenging for $m_\chi\lesssim$ 1 MeV. On the other hand, if $\chi$ is a subcomponent of the dark matter with $\Omega_\chi / \Omega_{\text{DM}}\lesssim 0.05$, the self-interaction constraints can be relaxed as discussed in Section~\ref{sec:selfint}. In this case we impose the conservative perturbativity bound $y_\chi<1$ for the right panel of Fig.~\ref{fig:scenA_money}. There is a subtlety associated with this regime: since $m_\phi < m_\chi$, such a large coupling implies that $\chi$ annihilates extremely efficiently to $\phi$, reducing its relic density to negligible levels. A straightforward way around this is to consider a complex scalar/Dirac fermion $\chi$ with an asymmetric relic abundance. In the regime of interest for subcomponent dark matter, the dark sector drops out of equilibrium well above the QCD phase transition, and the additional degrees of freedom required for asymmetric dark matter are allowed by the BBN bounds.

Both superfluid He and color centers could probe several orders of magnitude of new parameter space in this scenario. For the superfluid He projections shown in this paper, we have included the finite mediator mass and integrated over energies of 3 meV up to 100 eV in the nuclear recoil mode, leading to a slightly different reach compared to Ref.~\cite{Knapen:2016cue}.

For $m_\phi\gtrsim 500\,\keV$, we require that the dark matter $\chi$ does not thermalize with $\phi$, consistent with BBN bounds, and as explained in the previous section. We present the direct detection prospects in Fig.~\ref{fig:scenB_money} for two benchmarks, where we fix $y_\chi$ by saturating the $\phi$-$\chi$ thermalization and SIDM constraint, whichever is strongest. The various contours indicate the terrestrial and astrophysical constraints on $y_n$ discussed above. Since the self-interaction constraints are less stringent than the $\phi$-$\chi$ thermalization condition in this scenario, we do not find substantially different results for sub-component dark matter.
The turn-on of the $\phi\rightarrow \chi\chi$ decay mode at $m_\chi=m_\phi/2$ is clearly visible, and in practice we find that accessible cross sections are excluded whenever this decay is open. Then, if one allows for a somewhat large $\Delta \Neff \approx 0.57$, we find that there is available parameter space for $m_\chi>100$ keV with $y_n$ between the meson constraints and the supernova trapping window; these cross sections could be probed by experiments such as the nuclear recoil mode ($m_\chi>1$ MeV) in He, color centers, SuperCDMS and NEWS.

\section{Leptophilic scalar mediator \label{sec:scalarelectron}}

Analogous to the model in the previous section, here we take real scalar dark matter $\chi$ interacting with leptons via a scalar mediator. As long as the couplings of $\phi$ do not induce large lepton flavor-violation (\emph{e.g.}~with MFV couplings), the direct detection cross section and the bulk of the constraints depend only on the coupling to the electron. The effective Lagrangian is written as
\begin{align}
	\label{eq:electronmodel}
	\mathcal{L}\supset - \frac{1}{2}m_\chi^2\chi^2 - \frac{1}{2}m_\phi^2 \phi^2 - \frac{1}{2}y_\chi m_\chi \phi \chi^2  -y_e \phi \overline e e.
\end{align}
The discussion relating to vacuum stability of the scalar potential  and perturbativity  is identical to that in Section~\ref{sec:scalarnucleon} and Appendix~\ref{app:stability}, and we take $y_\chi<1$ everywhere. 
Again, we consider real scalar dark matter for simplicity, but our results also hold for fermion DM modulo important differences in the effects on BBN.
To account for the massive and massless mediator limits, we define a reference direct detection cross section,
\begin{align}
	\bar\sigma_e \equiv \frac{y_\chi^2  y_e^2}{4\pi}  \frac{ \mu_{\chi e}^2  }{(m_\phi^2 + \alpha^2 m_e^2)^2}
\end{align}
with $\mu_{\chi e}$ the DM-electron reduced mass. For electron scattering, the momentum transfer scale is set by the typical in-medium electron momentum $q \sim \alpha m_e$. As a result, here we define the light mediator limit by $m_\phi \ll \alpha m_e$. The leptophilic scalar model can be probed by superconductors~\cite{Hochberg:2015pha,Hochberg:2015fth}, Dirac materials~\cite{Hochberg:2017wce}, liquid xenon~\cite{Essig:2012yx,Essig:2017kqs}, graphene~\cite{Hochberg:2016ntt}, scintillators~\cite{Derenzo:2016fse}, and semiconductor detectors~\cite{Essig:2011nj,Essig:2015cda,Graham:2012su,Lee:2015qva}, among others~\cite{Battaglieri:2017aum}. 

\begin{figure}[t]
\centering
\includegraphics[width=0.75\textwidth]{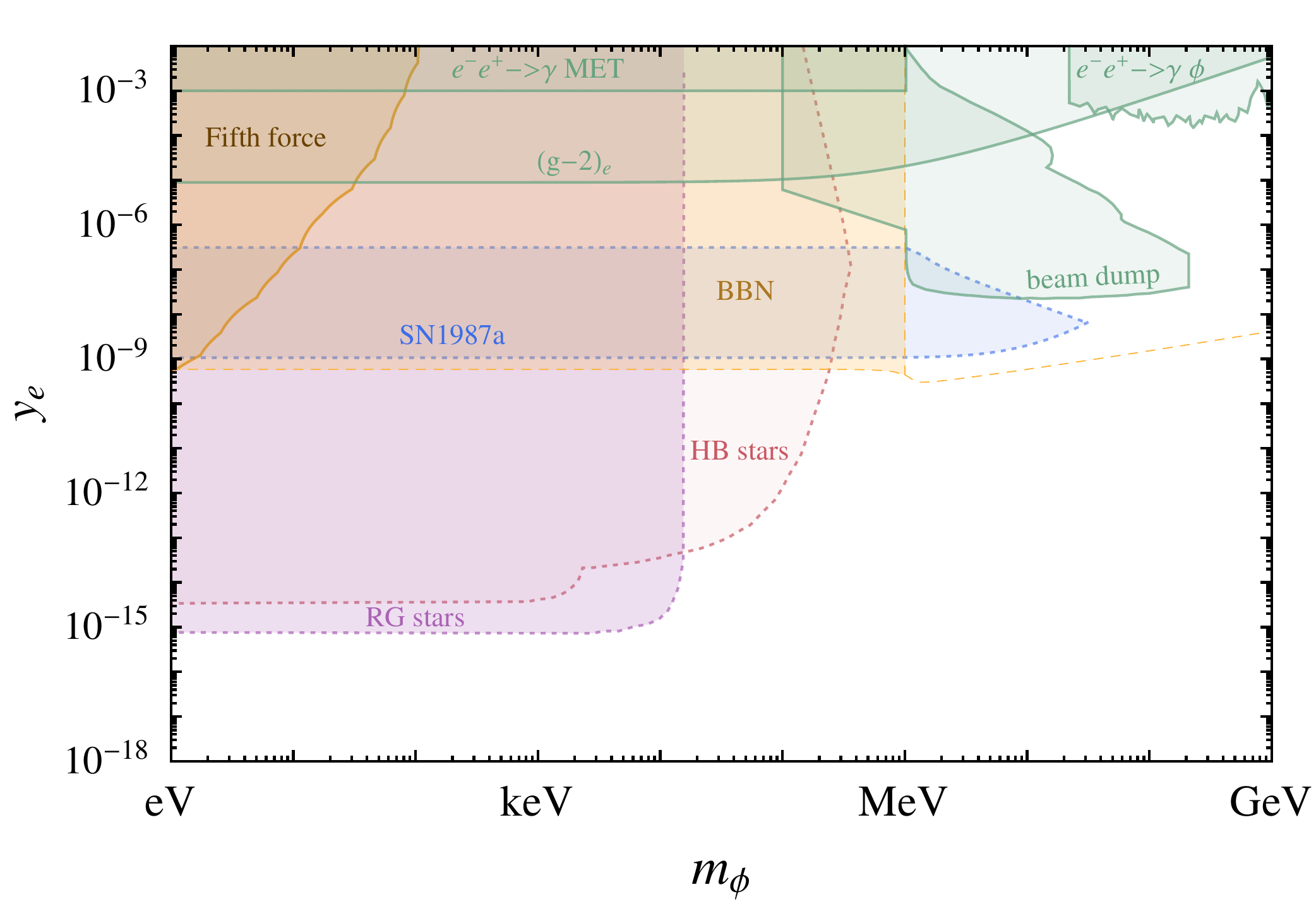} 
\caption{ Constraints on a sub-GeV scalar mediator, given in terms of the effective scalar-electron coupling $y_e$. We show terrestrial limits from fifth force searches \cite{Murata:2014nra} (orange), accelerator \cite{Essig:2013vha,Lees:2014xha,Liu:2016qwd} and $(g-2)_e$ constraints \cite{Liu:2016qwd,Liu:2016mqv} (green), and stellar cooling limits from HB stars \cite{Hardy:2016kme} (red), RG stars \cite{Hardy:2016kme} (purple) and SN1987A (blue). Note that the beam dump constraints, derived in Ref.~\cite{Liu:2016qwd}, assume only $\phi\to e^+e^-$ and $\phi \to \gamma \gamma$ decay modes are present, with negligible branching of $\phi$ to dark matter. Thermalization of $\phi$ before electron decoupling can occur for $y_e \gtrsim 10^{-9}$, indicated by the dashed yellow line. The shaded yellow region is excluded by BBN constraints, as discussed in the text. \label{fig:mphiplane_electron}}
\end{figure}

The constraints on the scalar mediator with electron coupling are shown in Fig.~\ref{fig:mphiplane_electron}, and described in more detail in the remainder of this section.  Note that these constraints have significant overlap with Higgs mixing models, which were considered recently in the context of thermal dark matter with $m_\chi>1$ MeV~\cite{Krnjaic:2015mbs}. The similarities are mostly in the very stringent stellar constraints on electron couplings of light scalars, which in Higgs mixing models dominate over nucleon couplings despite the Yukawa coupling suppression. Important differences arise in the context of terrestrial constraints, where the absence of a coupling to hadrons lifts some of the accelerator and meson decay bounds.
Cosmologically, the most important difference compared to the hadrophilic scalar is that here $\phi$ can enter into equilibrium as the universe cools, and furthermore can remain in equilibrium through $T<m_\pi$. This results in more robust BBN constraints on mediators with mass below a few MeV. We describe the cosmology further in Section~\ref{sec:cosmo_electron}.

\subsection{Terrestrial constraints}

With only a coupling to electrons, the dominant terrestrial constraints on the mediator are derived from precision measurements of $(g-2)_e$ \cite{Liu:2016qwd,Liu:2016mqv}, and from $\phi$ direct production in high intensity $e^+ e^-$ colliders or beam dump experiments. For $m_\phi <2m_e$, $\phi$ must decay to dark matter or photons, where the latter decay is suppressed by $\alpha^2/16\pi^2$.  If $\phi$ either decays invisibly or outside the detector, we can apply the BaBar mono-photon limits from Ref.~\cite{Essig:2013vha}. While this constraint is one of the more robust limits in Fig.~\ref{fig:mphiplane_electron}, it is not competitive with the BBN and stellar constraints, which we discuss below. When $m_\phi \gtrsim 20 \mbox{ MeV}$ the BaBar dark photon search for  $e^+e^-\to \gamma  (\phi\to \ell^-\ell^+)$ \cite{Lees:2014xha} can be used to set a constraint, provided that $\phi$ couples to muons and electrons with mass-hierarchical couplings.  In particular, while the dark photon model considered in Ref.~\cite{Lees:2014xha} has democratic branching ratios between muons and electrons, most of its sensitivity comes from the muon channel. For $\text{Br}[\phi \to \mu\mu]\gg \text{Br}[\phi \to ee]$, a limit on $\phi$ can be approximated using the limit on the dark photon model considered by BaBar. We rescale the limit to account for the hadronic branching fraction in the dark photon model \cite{Batell:2009yf}, which is absent in the model we consider here. If $\phi$ has a small or zero branching ratio to muons, the limit is weaker by an order one factor. A recasting of Ref.~\cite{Lees:2014xha} is needed in this case, which we do not attempt here. 

In the mass range $m_\phi \gtrsim 100$ keV, electron beam dump experiments provide stringent constraints \cite{PhysRevD.38.3375,PhysRevLett.59.755,DAVIER1989150}, as derived for the $\phi \bar e e $ coupling in Ref.~\cite{Liu:2016qwd}. The beam dump constraints may be relaxed if the visible decays of $\phi$ are suppressed by a competing decay mode to invisible states. This is especially likely to occur for $m_\phi<2m_e$, where the dominant visible mode is the radiative decay to photons.  Finally, fifth force constraints become important for mediator masses below an eV; they are, however, weaker by a factor $m_e/m_n$ compared to the hadrophilic scalar, and so are not competitive with stellar and BBN bounds in the mass range we consider. There are also bounds on light scalars from measurements of splittings in positronium~\cite{Delaunay:2017dku}, although these are currently weaker than the  bound from $(g-2)_e$.

\subsection{Astrophysical constraints \label{sec:stellar_electron}}

The self-interaction constraints on this model are identical to those for the hadrophilic scalar, and we refer the reader to Section~\ref{sec:selfint}.  The stellar constraints on the other hand differ quantitatively, as the rate for producing a light mediator off an electron is enhanced compared to the rate off a comparatively heavier nucleon. The strongest bounds are however still obtained from horizontal branch (HB) and red giant (RG) stars for $m_\phi \lesssim 100 $ keV and from SN1987A for heavier mediators.  We take the HB and RG limits from Ref.~\cite{Hardy:2016kme}, which account for plasma mixing effects, and correspond to $g_e \lesssim 7\times10^{-16}$ in the massless limit. Ref.~\cite{Hardy:2016kme} also included the effects of trapping for couplings as large $ g_e \approx 10^{-6}$, which weakens the bounds somewhat in these stars. We show the HB and RG bounds in Fig.~\ref{fig:mphiplane_electron}, where we have extrapolated their results to even larger couplings (as this region is separately excluded by BBN bounds).

Complete constraints from SN1987A have not yet been derived for this model. We estimate these bounds in the limit that only production via mixing with the longitudinal component of the photon is included, and neglecting direct production via Compton scattering or electrion-ion interactions. As shown in Ref.~\cite{Hardy:2016kme}, this is a ``resonant'' production because the energy of the emitted scalars is $\omega = \omega_L$, with $\omega_L$ the frequency where the scalar and longitudinal photon dispersions cross. From Ref.~\cite{Hardy:2016kme} (see Appendix A.4), the energy loss rate from resonant production is given by
\begin{align}
	Q_{\rm res} \simeq \frac{\omega_L}{4\pi} \left( \frac{\omega_L}{m_\phi} \Pi^{\phi L} \right)^2  \frac{1}{e^{\omega_L/T} - 1},
\end{align}
where we have used the fully relativistic result. $\Pi^{\phi L}$ is the mixing of the scalar with the longitudinal component of the photon in the medium, which can be written as
\begin{align}
	\Pi^{\phi L} \simeq \frac{y_{e} e \, m_e^{\rm eff} m_\phi}{ \pi^2 k} \int_0^\infty dp \, v^2 \, \left(f_e(E_p) + f_{\bar e}(E_p) \right) \left( \frac{\omega_L}{v k} \log \left( \frac{\omega_L + v k}{\omega_L - v k} \right) - \frac{2 m_\phi^2}{ \omega_L^2 - k^2 v^2} \right),
\end{align}
where $f_e$ and $f_{\bar e}$ are the phase space distributions for the electrons, $v = p/E_p$ is the electron momentum, and $k = \sqrt{\omega_L^2 - m_\phi^2}$ is the 3-momentum of the mediator $\phi$. Note that the result is proportional to the in-medium mass of the electron, $m_e^{\rm eff} \approx 12$ MeV in the core of the supernova.
Using the Raffelt condition on the energy loss per unit mass $\epsilon = Q/\rho \lesssim 10^{19}$ erg/g/s with $T \approx 30$ MeV, $\omega_L \approx 82$ MeV, and $\rho \approx 3 \times 10^{14}$ g/cm$^3$, we obtain a limit for the weak coupling regime of $y_e \lesssim 10^{-9}$ for massless scalars. Given that we have only included resonant production, we expect that the true bounds due to thermal production may be even stronger.

To derive the trapping regime for SN1987a, we again use detailed balance to relate the production rate to the absorption rate. For resonant production, $\lambda_{\rm mfp}^{-1} = \Gamma_{\rm abs}(\omega) \sim Q_{\rm res}/\omega_L^4$. Requiring that the scalar is re-absorbed within $R \approx 10 $ km leads to a trapping limit of $y_e >  3 \times 10^{-7}$.  We also account for trapping due to the decay of $\phi \to e^+ e^-$, where we require that the decay length of $\phi \to e^+ e^-$ is within $R \approx 10$ km. The decay of $\phi$ determines the bound in the trapping regime for masses MeV $\leq m_\phi \leq$ 30 MeV. (Note that in computing kinematically allowed decays to $e^+e^-$, we use the vacuum mass $m_e$ as opposed to the effective mass $m_e^{\rm eff}$. This is because the thermal corrections to the electron mass drop rapidly and become smaller than the bare $m_e$ for $R$ beyond a few km, depending on the model assumed. Hence, while we use the in-medium $m_e^{\rm eff}$ for production in the core, we simply use $m_e = 511$ keV to calculate decay.) 

\subsection{Cosmology \label{sec:cosmo_electron}}

For large enough couplings, the mediator $\phi$ will be in thermal contact with the standard model through annihilation ($e^+e^-\rightarrow \gamma \phi$) and Compton scattering ($e^-\gamma\rightarrow e^-\phi$).  The decays $\phi\rightarrow e^+ e^-$ (if $m_\phi>2m_e$) and $\phi \rightarrow \gamma\gamma$ (if $m_\phi<2m_e$) also contribute to equilibrating $\phi$ with the standard model. Both the scattering and decay processes are IR dominated, as compared to the Hubble expansion, such that the mediator enters thermal equilibrium as the universe cools. This is qualitatively different from the nucleon coupling model, where the coupling with the standard model was provided by the UV dominant, dimension-five $\phi GG$ operator. In practice we find that decays are always subdominant to the Compton and annihilation processes, regardless of $m_\phi$. In the limit $s\gg m_\phi^2, m_e^2$, the cross sections are
\begin{align}
\sigma_{e\gamma \to e \phi}&\approx\frac{\alpha y_e^2}{s}\left[\log\left(\frac{s}{m_e^2+m_\phi^2}\right)+\frac{5}{2}\right]\\
\sigma_{ee \to \gamma \phi}&\approx\frac{2\alpha y_e^2}{s}\log\left(\frac{s}{4m_e^2}\right).
\end{align}
The thermally averaged cross section for annihilation is obtained by replacing $m_\phi\rightarrow m_e$ in Eq.~\eqref{eq:thermaver}, while the corresponding formula for Compton scattering is
\begin{equation}
	\langle \sigma_{e\gamma \to e \phi} v \rangle=\frac{1}{16 {m^2_e} T^3 K_2(m_e/T)}\int^{\infty}_{m_e^2}\!\!ds\, \sigma \,(s-m_e^2)\sqrt{s} K_1(\sqrt{s}/T).
\end{equation}
Analogous to the discussion in Section~\ref{sec:cosmo}, we say the mediator thermalizes if the thermally averaged rate is greater than the Hubble expansion $H(T)$ at $T\approx \mathrm{max}[1\,\MeV , m_\phi]$. This yields the dashed yellow line in Fig.~\ref{fig:mphiplane_electron}; below this line, the mediator does not come in thermal contact with the standard model while electrons and the mediator are both still relativistic. For  $m_\phi\ll 1\,\MeV$, this value is $y_e\lesssim 5\times 10^{-10}$ and independent of $m_\phi$.  Above the electron threshold, the mediator decouples at $T\sim m_\phi$ and the bound on $y_e$ therefore scales as $\sim \sqrt{m_\phi/M_{pl}}$.

\begin{figure}[t]
\hspace{-0.2cm}
\includegraphics[width=0.49\textwidth]{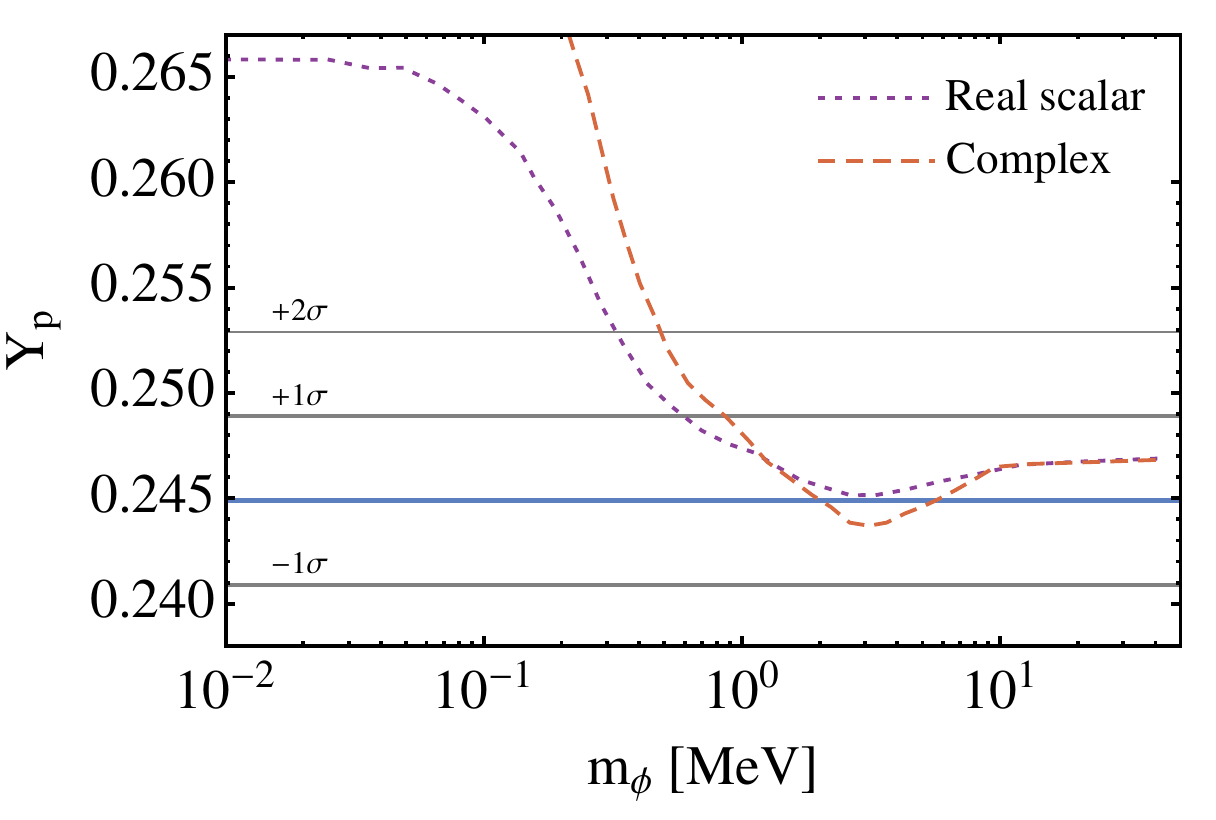}\hfill
\includegraphics[width=0.48\textwidth]{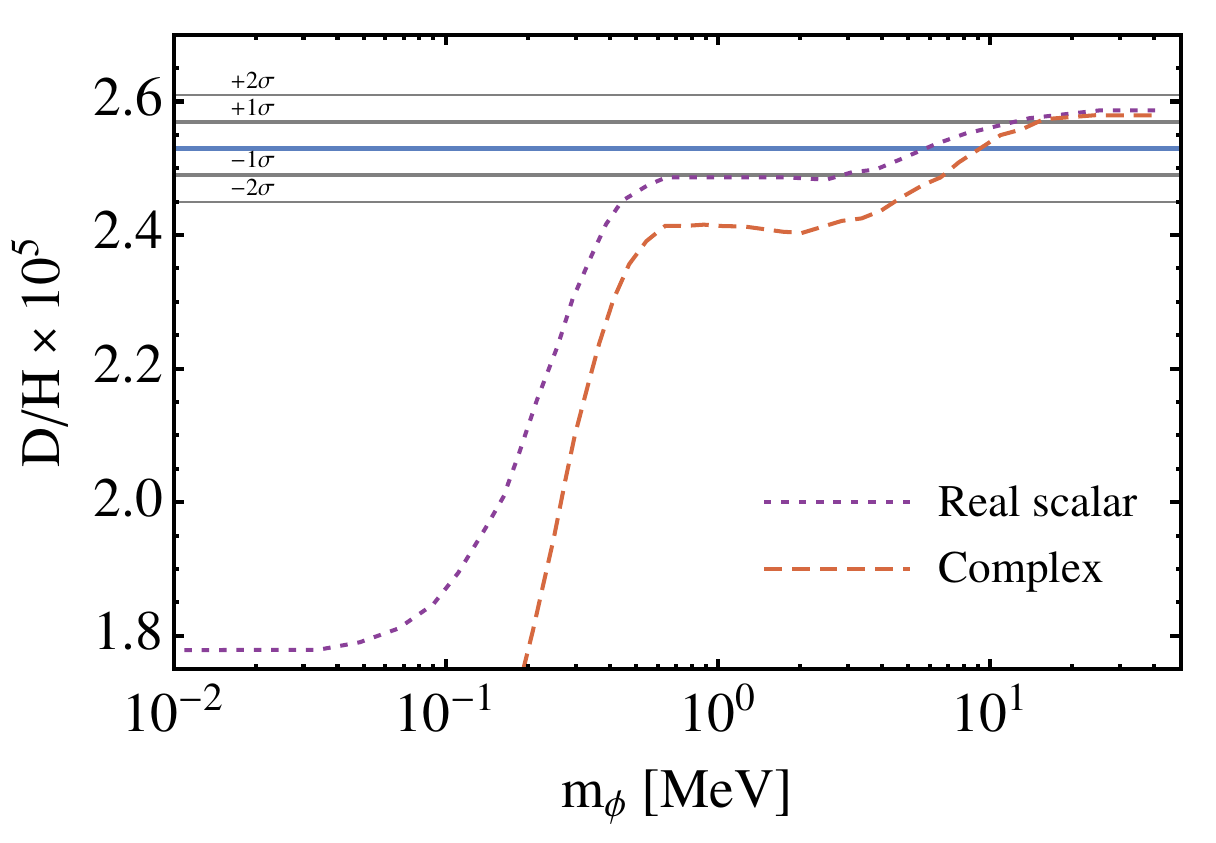}
\caption{ For a real (complex) scalar degree of freedom in equilibrium with electrons/photons, we compare the results of Ref.~\cite{Boehm:2013jpa} with the updated measurements of the BBN abundances of helium ($Y_p$) and deuterium ($D/H$) from Ref.~\cite{Cyburt:2015mya,Cooke:2013cba}. We show the central values of the measurements (solid blue lines) along with $1\sigma$ and $2\sigma$ bands.
\label{fig:celine}
}
\end{figure}

For  $y_e\gtrsim 5\times 10^{-10}$, $\phi$ can enter equilibrium with electrons before $T \sim $ MeV, potentially running afoul of BBN. In particular, any light degrees of freedom in thermal equilibrium with $\gamma/e$ will decrease the deuterium abundance~\cite{Boehm:2013jpa}. Here, the presence of $\phi$ dilutes the entropy release from $e^+e^-$ annihilation at the electron mass threshold, which has a bigger effect than increasing $\Neff$ alone. This is because the photon temperature $T_\gamma/T_\nu$ is reduced during BBN, leading to a increased baryon-to-photon ratio $\eta$ and thus a more efficient conversion of deuterium into He. We compare the results of Ref.~\cite{Boehm:2013jpa} with new measurements of the helium fraction $Y_p$ and of D/H~\cite{Cyburt:2015mya,Cooke:2013cba} in Fig.~\ref{fig:celine}. It follows that $m_\phi$ below $m_e$ is in tension with current measurements, regardless of whether $\phi$ is in equilibrium with other dark degrees of freedom. This is the meaning of the yellow shaded region in Fig.~\ref{fig:mphiplane_electron}.

The bounds further strengthen if $\phi$ and $\chi$ are both in equilibrium with the SM. Similar to the discussion above, Fig.~\ref{fig:celine} indicates that two real scalars with mass below $\approx 5 \mbox{ MeV}$ are in tension with the new deuterium measurements. 
For this statement, we used the complex scalar benchmark from \cite{Boehm:2013jpa}, implicitly assuming $m_\chi\approx m_\phi$, as well as the absence of other dark degrees of freedom in the thermal bath. (The case for a non-degenerate $\phi$ and $\chi$ would require a dedicated study, which we do not attempt here.) We  note that for $m_\phi\gtrsim 10$ MeV, $\phi$ could in principle transfer its entropy back to the SM thermal bath before neutrino decoupling, for instance by freezing out against the electrons. In this case the $m_\chi \gtrsim m_e$ bound should be sufficient, even if $\chi$ remains in equilibrium with the electrons through off-shell $\phi$ exchange. This is an important difference with the hadrophilic scalar in Section~\ref{sec:scalarnucleon}: once the temperature drops below $m_\pi$ in the nucleon coupling model, $\phi$ and/or $\chi$ have no more means to effectively communicate with the SM, and all their entropy must be dumped into other dark sector degrees of freedom. This increases $\Delta \Neff$, even if $m_{\chi,\phi}\gg1$ MeV.  

For $m_\chi\lesssim 1$ MeV, other dark sector states must be present in order to set the $\chi$ relic abundance, similar to the interactions in Eq.~\eqref{eq:lightdof}. Clearly, the precise constraints from BBN depend on the details of how the relic abundance of $\chi$ is set and a dedicated study is necessary to map out the full allowed parameter space. In what follows, we will restrict ourselves instead to two conservative benchmark points as far as the heavy mediator regime is concerned: 
\begin{itemize}
\item Fix $m_\phi = 10 \mbox{ MeV}$, and allow $\chi$ to thermalize with $\phi$ as long as  $m_\chi > m_e$.
\item Fix $m_\phi = m_\chi> 5$ MeV, and allow $\chi$ to thermalize with $\phi$, such that the bound for a complex scalar in Fig.~\ref{fig:celine} is satisfied.
\end{itemize}

\subsection{Results}

\begin{figure}[th]
\includegraphics[width=0.49\textwidth]{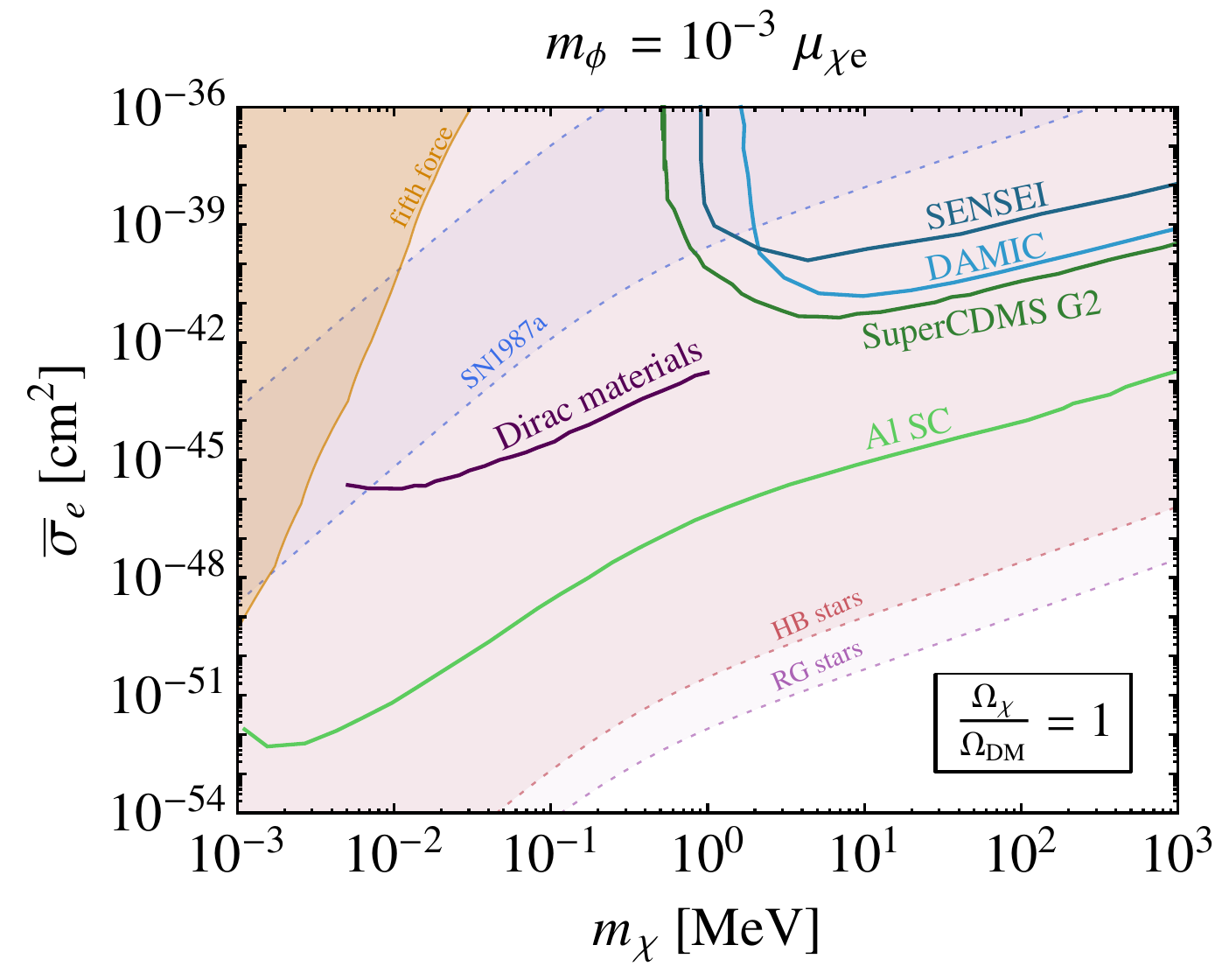}\hfill
\includegraphics[width=0.49\textwidth]{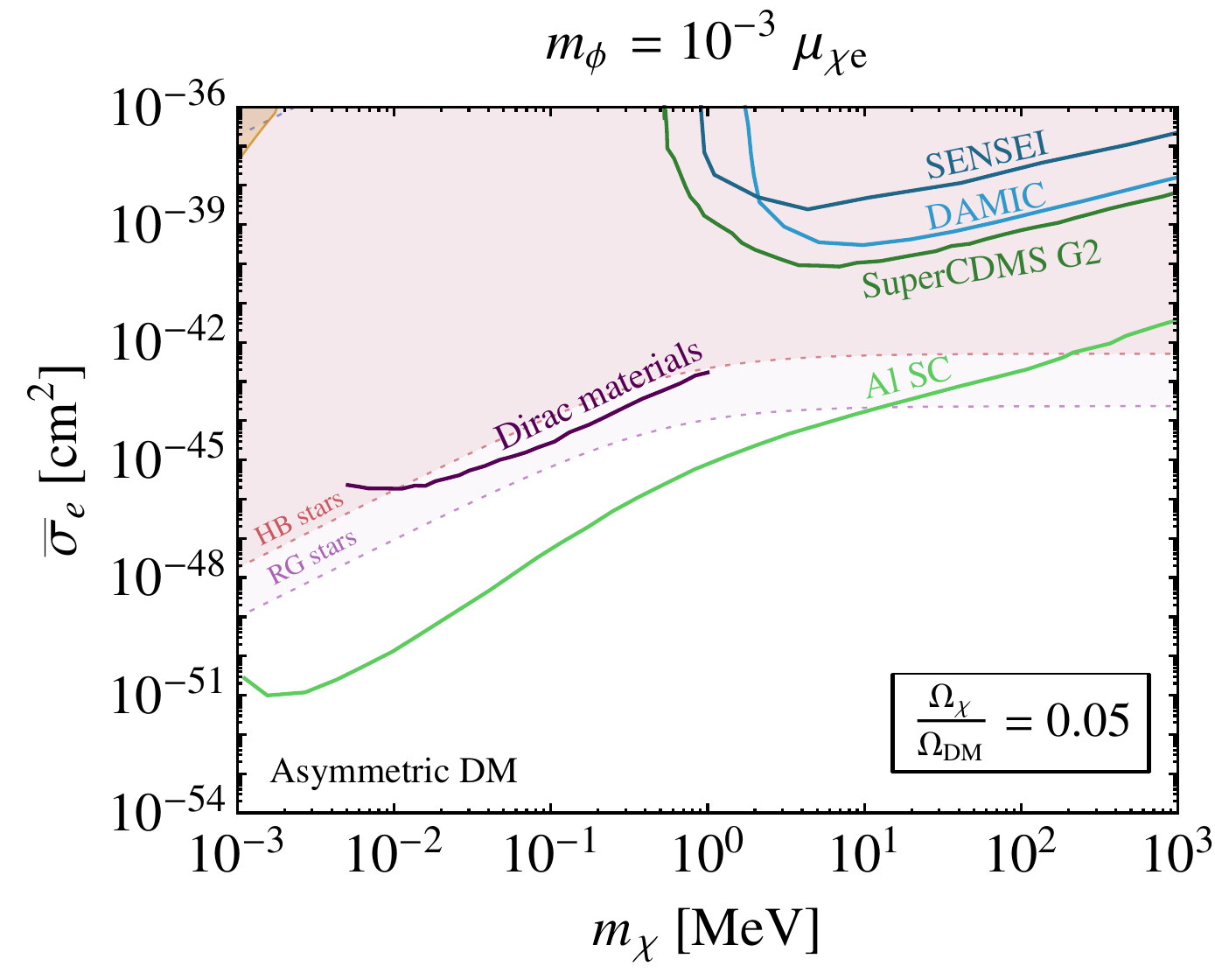}
\caption{Direct detection cross section as function of the dark matter mass for the case that $\chi$ is all the dark matter (left) and when $\chi$ composes $5\%$ of the dark matter (right).  In the former case $y_\chi$ is fixed by saturating the self-interaction constraint, while in the latter case we take $y_\chi=1$  and assume $\chi$ is a complex scalar with an asymmetric abundance. We take $m_\phi=10^{-3} \mu_{\chi e}$, such that we are in the light mediator limit. The various lines indicate the reach for SuperCDMS-G2+, SENSEI-100g, DAMIC-1K \cite{Battaglieri:2017aum}, an aluminum superconducting (Al SC) target \cite{Hochberg:2015fth}, or a Dirac material \cite{Hochberg:2017wce}, assuming kg-year exposure in all cases. See Ref.~\cite{Battaglieri:2017aum} for other proposals that could probe this parameter space.  Current bounds from Xenon10 and Xenon100 are present only for cross sections above the range shown. \label{fig:scenA_money_phiee}}
\end{figure}

\begin{figure}[t]
\includegraphics[width=0.49\textwidth]{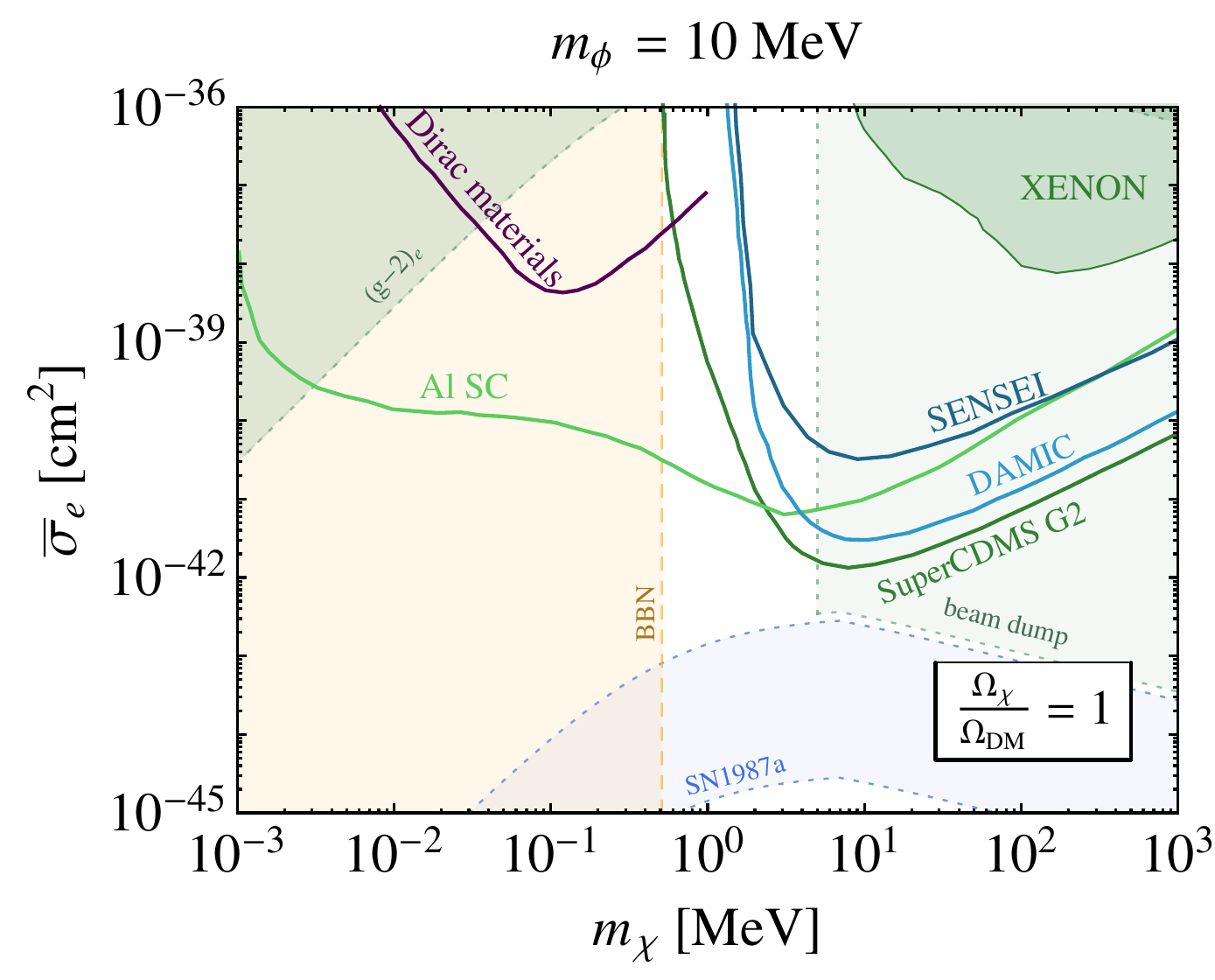}\hfill
\includegraphics[width=0.49\textwidth]{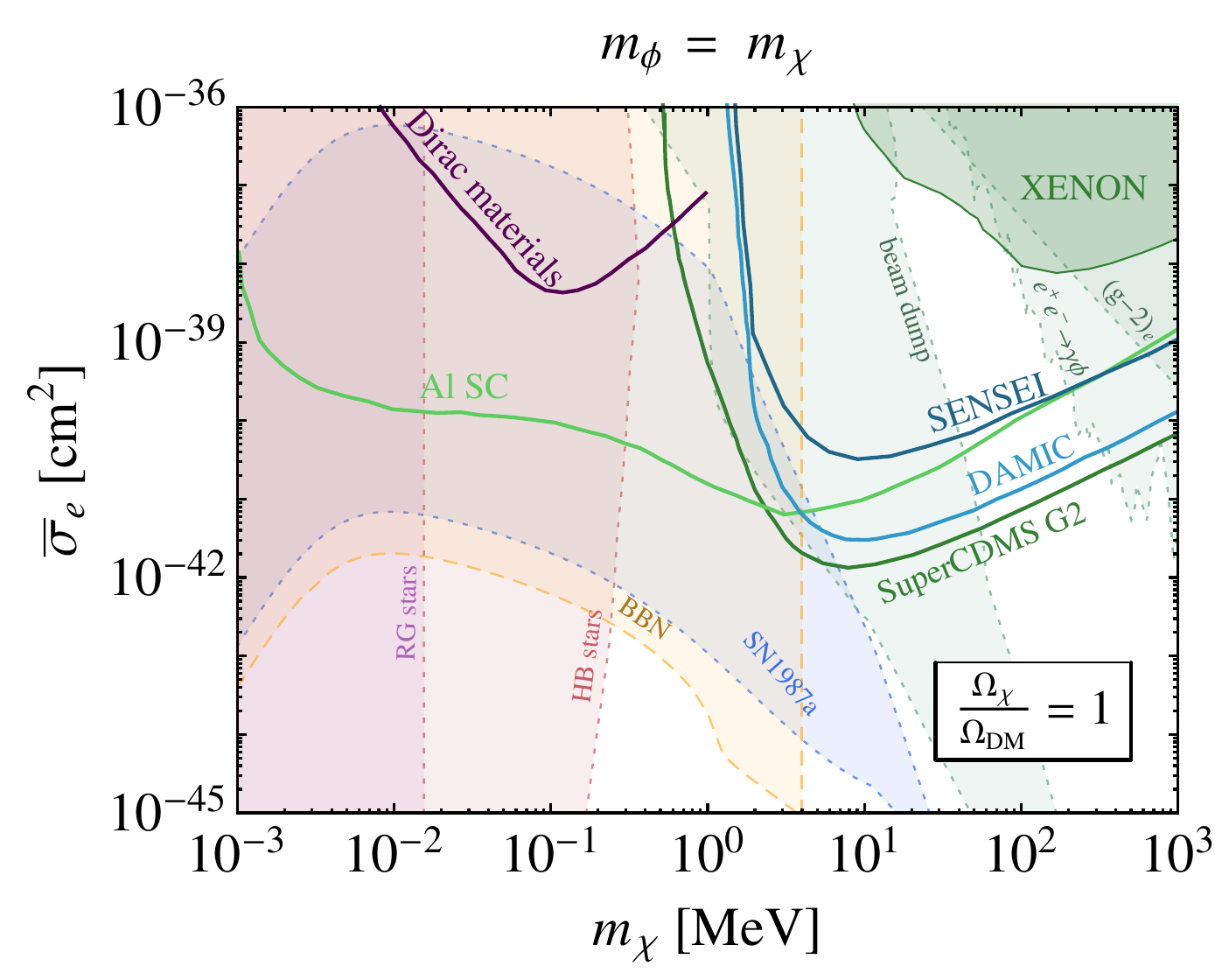}
\caption{Direct detection cross section as function of the dark matter mass for $\chi$ being all the dark matter, for two different massive mediator benchmarks. $y_\chi$ is fixed by saturating the self-interaction constraint.  The various lines indicate the reach for SuperCDMS-G2+, SENSEI-100g, DAMIC-1K \cite{Battaglieri:2017aum}, an aluminum superconducting (Al SC) target \cite{Hochberg:2015fth}, or a Dirac material \cite{Hochberg:2017wce}, assuming kg-year exposure in all cases. See Ref.~\cite{Battaglieri:2017aum} for other proposals that could probe this parameter space. Current bounds from Xenon10 and Xenon100 are also shown~\cite{Essig:2012yx,Essig:2017kqs}. 
Note that the constraint from beam dumps in the right panel (for masses of MeV to 100 MeV) can be lifted if $\phi$ decays to invisible states.
\label{fig:scenB_money_phiee}
}
\end{figure}

Motivated by the combination of cosmological, astrophysical, and terrestrial constraints discussed above,  we consider two regimes (similar to the hadrophilic scalar) which are demarcated according to whether the mediator is very light or massive:
\begin{itemize}
\item  $m_\phi\ll m_\chi$  and $y_e\lesssim 10^{-15}$, where the bound on $y_e$ arises from stellar constraints. The dark sector is never in equilibrium with the standard model and SIDM constraints provide the strongest bounds on $y_\chi$ (see Fig.~\ref{fig:alphaXtherm}), if $\chi$ is the dominant component of the dark matter. The available parameter space in this scenario is shown in the left hand panel of Fig.~\ref{fig:scenA_money_phiee}, where we included the expected DAMIC and SuperCDMS G2+ reach as representative examples for semiconductor targets, as well as the reach for Dirac materials and superconductors. We find that this scenario is not accessible with existing proposals, in agreement with earlier findings \cite{Hochberg:2015fth}.

We also show the case where $\chi$ is 5\% of the total dark matter density, in the right hand panel of Fig.~\ref{fig:scenA_money_phiee}, assuming that SIDM constraints are lifted and setting $y_\chi=1$. Similar to the hadrophilic model, here we take $\chi$ to be a complex scalar with an asymmetric relic abundance. The symmetric component of the $\chi$ abundance is then rapidly depleted through $\chi\bar\chi\to\phi\phi$ annihilation. The $\phi \bar e e$ coupling remains challenging even in this case, and only superconductors are expected to have sensitivity. It is however conceivable that the stellar constraints could be lifted or weakened in a more sophisticated model, and that reach for Dirac materials or semiconductor targets could be recovered.

\item $m_{\phi} \gtrsim 10\, \MeV$ and $m_\chi \gtrsim m_e$, and $m_\phi = m_\chi> 5$, both with $y_e \gtrsim  5\times10^{-10}$: For these couplings, the mediator and the dark matter are in equilibrium with the SM in the early universe, but disappear from the thermal bath sufficiently early to satisfy BBN constraints (in particular from the deuterium abundance). Fig.~\ref{fig:scenB_money_phiee} shows the resulting parameter space in this massive mediator scenario. In the left panel, we take $m_\phi = 10$ MeV, motivated by the window between the beam dump and $(g-2)_e$ constraints in Fig.~\ref{fig:mphiplane_electron}. There is detectable and potentially unconstrained parameter space when the DM is heavier than the electron mass, although further studies of the BBN predictions in this case are required. In this part of parameter space one finds $y_e\ll y_\chi$, such that the beam dump constraints do not apply if the $\phi\rightarrow\chi\chi$ decay mode is kinematically accessible ($m_\chi<5$ MeV in the left hand panel of Fig.~\ref{fig:scenB_money_phiee}). 

In the right panel of Fig.~\ref{fig:scenB_money_phiee}, we instead fix $m_\phi=m_\chi$ and exclude the $m_\phi = m_\chi < 5$ MeV from BBN considerations (yellow region), as discussed in the previous section. The terrestrial bounds allow fairly large $\bar \sigma_e$ for $m_\chi > 10\, \MeV$. Note also that we have conservatively applied the beam dump constraints, which excludes a range of cross sections in the mass range from 1~MeV to 100~MeV -- however, these could be lifted if $\phi$ can decay to invisible states (either by taking $m_\phi > 2 m_\chi$ or by introducing other dark sector states).

\end{itemize}

\clearpage

\section{Vector mediators \label{sec:otherscenarios}}

We now discuss vector mediators, concentrating on the simplest two anomaly-free extensions to the SM: a kinetically mixed dark photon and a $U(1)_{B-L}$ gauge boson. For these benchmark models we assume Dirac dark matter, and as such the self-interaction constraints are slightly different from the real scalar dark matter in the previous sections. We review this in Appendix~\ref{app:SIDM}.
In the scalar mediator scenarios discussed in Section~\ref{sec:scalarnucleon} and \ref{sec:scalarelectron}, constraints were broadly characterized by a ``light mediator'' regime and a ``massive mediator'' regime, with $m_\phi \sim$ a few hundred keV as the rough boundary between the two.  In the light mediator regime, the constraints are driven by stellar cooling and fifth force bounds. In the massive mediator regime, cross sections were instead limited due to terrestrial bounds, and $\Neff$ bounds on thermalization of the dark sector (with $\Neff$ primarily important for sub-MeV dark matter). 

The vector mediator cases differ notably from the scalar mediated models in that (i) stellar constraints on the mediator decouple in the massless limit and (ii) the BBN bounds on the massive mediator scenario are even more stringent than those of Section~\ref{sec:cosmo_electron}, given the larger number of degrees of freedom. 
Due to the BBN bounds, realistic cross sections for $m_\chi \lesssim$ MeV are difficult to obtain with a massive mediator. Moreover, for $m_\chi\gtrsim$ MeV massive vector mediator models have been discussed extensively in the literature already~\cite{Lin:2011gj,Chu:2011be,Izaguirre:2014bca,Izaguirre:2015yja,Izaguirre:2015zva,Izaguirre:2017bqb}, especially for the case of a kinetically mixed dark photon. For these reasons, we will focus exclusively on the available parameter space in the light mediator regime.

\subsection{Kinetically mixed dark photon}

The interactions for this model are given by
\begin{align}\label{eq:darkphotonlag}
	{\cal L} \supset = - \frac{1}{2} m_{A'}^2 A^\prime_\mu A'^\mu -\frac{1}{4} F'^{\mu \nu} F^\prime_{\mu \nu} -\frac{\epsilon}{2} F^{\mu \nu} F^\prime_{\mu \nu} - y_\chi A^\prime_\mu \bar \chi \gamma^\mu \chi \ ,
\end{align}
where now we consider Dirac fermion dark matter. For simplicity, we assume that the mass of the dark vector was generated by the Stueckelberg mechanism. The kinetic mixing, parameterized by $\epsilon$, gives rise to a coupling of the dark photon with electrons and protons. From this, we define the reference electron-scattering cross section as
\begin{align}
	\bar \sigma_e \equiv  \frac{4 y_\chi^2  \alpha \epsilon^2   \mu_{\chi e}^2  }{(m_{A'}^2 + \alpha^2 m_e^2)^2}.
\end{align}
The scattering form factor for a light $A'$ is given by $F^2(q^2)= \alpha^4 m_e^4/q^4$. 
\begin{figure}
\centering
\includegraphics[width=0.6\textwidth]{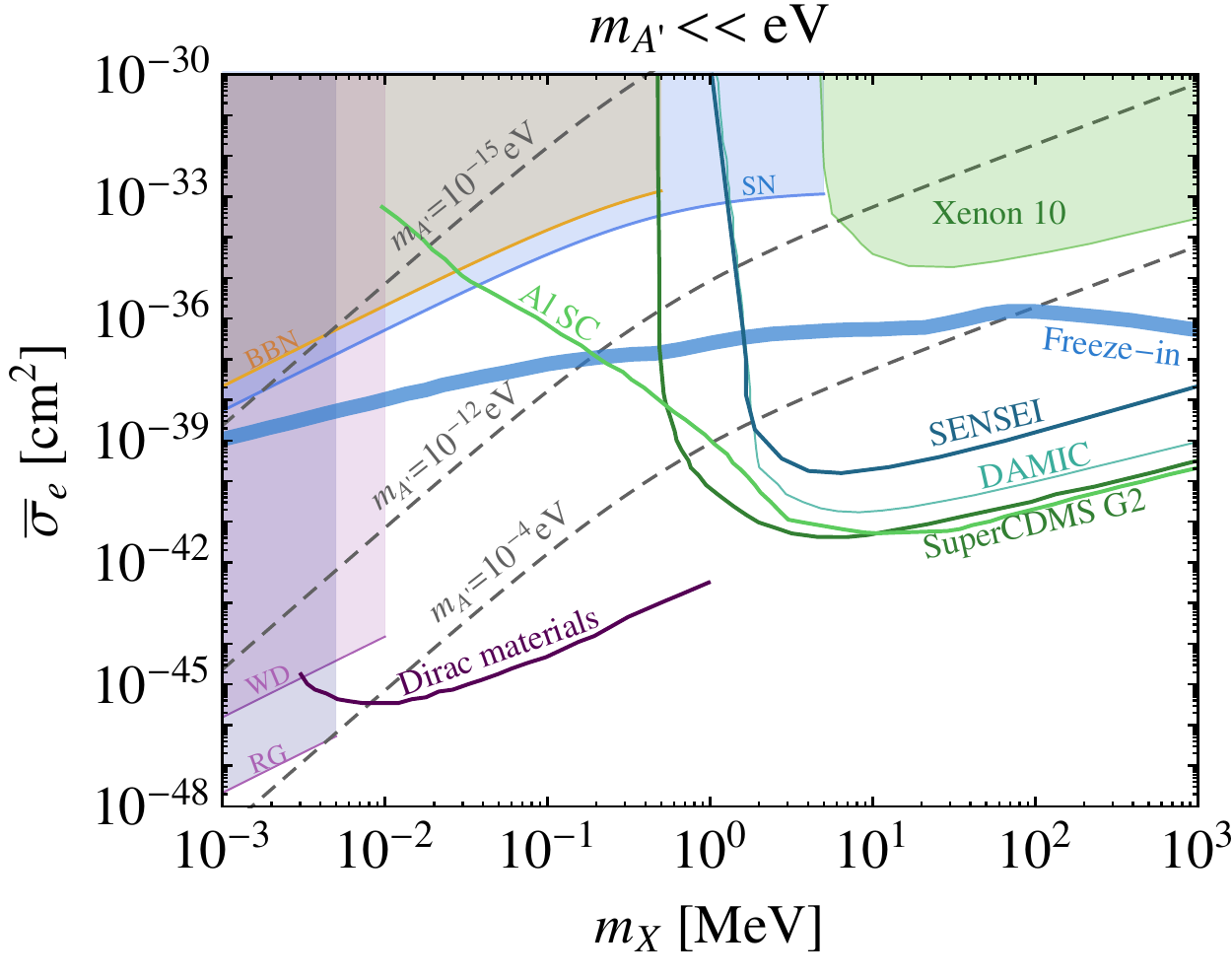} 
\caption{Dark matter scattering via a kinetically mixed dark photon, in the limit where $m_{A'} \ll $ keV. The shaded regions show millicharged DM limits (from BBN, SN1987A, and WD/RG)~\cite{Davidson:2000hf} as well as Xenon10 bounds~\cite{Essig:2012yx}. 
The dashed lines are upper bounds on the cross section, for several benchmarks where we have considered roughly the largest allowable $\epsilon$: (a) $m_{A'} = 10^{-15}$ eV, $\epsilon = 10^{-3}$, where stellar and fifth force constraints have decoupled (b) $m_{A'} = 10^{-12}$ eV, $\epsilon = 10^{-6}$, consistent with the CMB bounds shown in \cite{Mirizzi:2009iz}, and  (c) $m_{A'} = 10^{-4}$~eV, $\epsilon = 10^{-8}$, consistent with the stellar bounds in \cite{An:2013yua}. Here the DM-mediator coupling $y_\chi$ is fixed by saturating SIDM constraints. For larger $m_{A'}$ (up to 100 keV), the bounds on $\epsilon$ are much stronger.  Similar to Fig.~\ref{fig:scenB_money_phiee}, we also show the reach for various direct detection proposals.
\label{fig:darkphoton}}
\end{figure}

Unlike constraints on scalar mediators, when a vector mediator couples to SM particles proportional to electric charge, the stellar constraints decouple due to in-medium effects as $m_{A'} \to 0$. We review the derivation of this effect in Appendix~\ref{app:mixing}. For dark photons, these in-medium effects have been accounted for in the Sun, red giants, and horizontal branch stars in Refs.~\cite{An:2013yfc,An:2013yua,An:2014twa}, while the SN1987A limits have recently been updated in Refs.~\cite{Chang:2016ntp,Hardy:2016kme}.
In all cases, the limits on the kinetic mixing parameter $\epsilon$ scale as $1/m_{A'}$. The in-medium suppression of the $A'$ coupling with the electromagnetic current also implies thermalization of $A'$ is a negligible effect, thus avoiding cosmological bounds on $A'$. 

These bounds are however for direct emission of the $A'$, but light dark matter can also be emitted via an off-shell photon in the medium. As reviewed in Appendix~\ref{app:mixing}, this process does {\emph{not}} decouple in the massless limit. One way to see this is that for $m_{A'}\rightarrow 0$, $\chi$ is effectively millicharged with respect to the SM photon in the stellar medium. Hence the stellar and BBN constraints on millicharged particles can be applied to the DM \cite{Davidson:2000hf} and are directly sensitive to the combination  $\epsilon y_\chi/e$, which is the effective millicharge of the DM in the $m_{A'} \to 0$ limit. In other words, in the massless $A'$ limit, both the direct detection cross section as well as the stellar and BBN constraints become independent of $m_{A'}$. We can therefore map the millicharge constraints directly in the $m_\chi$ vs $\bar \sigma_e$ plane, indicated by the shaded regions in Fig.~\ref{fig:darkphoton}. Furthermore, $\chi$ gains its relic abundance through production via an off-shell photon, as in the stellar medium. The solid blue line in Fig.~\ref{fig:darkphoton} labels couplings where the correct abundance can be achieved through freeze-in~\cite{Essig:2011nj,Chu:2011be,Essig:2015cda}.

Even though the stellar constraints decouple for $m_{A'}\ll$ eV, astrophysical constraints and tests of deviations from Coulomb's law still constrain the mixing parameter $\epsilon$ as a function of $m_{A'}$. For a summary, see for example Refs.~\cite{Jaeckel:2010ni,Essig:2013lka,An:2013yua}. (The fifth force constraints we have considered previously are for macroscopic neutral systems, and are generally not relevant for the kinetically mixed dark photon.) For reference, we have included a number of benchmark lines in Fig.~\ref{fig:darkphoton}: for each $m_{A'}$ we have selected the largest allowed $\epsilon$ and we have fixed $y_\chi$ by saturating SIDM constraints, such that direct detection cross sections above the dashed line are excluded for the corresponding value of $m_{A'}$. We see that for sub-MeV dark matter, ultralight mediators are required to satisfy both SIDM constraints and for $\chi$ to live on the freeze-in line.

\subsection{$B-L$ gauge boson \label{sec:B_L}}

Next we consider gauging the $U(1)_{B-L}$ symmetry of the standard model, with gauge coupling $g_{B-L}$. To ensure the model is anomaly free, it suffices to consider Dirac neutrinos, or to add a set of 3 (heavy) sterile neutrinos. In order to avoid complicating the cosmology, here we follow the latter avenue, which implies that $U(1)_{B-L}$ is broken softly by the Majorana neutrino masses at a scale $\sim m_{A'}/{g_{B-L}}$.  For most of the parameter space of interest, the sterile neutrinos can be as heavy as a few GeV. The relevant constraints for $U(1)_{B-L}$ are summarized in Fig.~\ref{fig:BLplane}, while the alternative case with an unbroken $U(1)_{B-L}$ is discussed in Ref.~\cite{Heeck:2014zfa}. For the sake of generality, we allow for different values for the coupling of $A'$ to the SM $U(1)_{B-L}$ current ($g_{B-L}$) and to the dark matter ($y_\chi$, as in Eq.~\eqref{eq:darkphotonlag}). We note that in parts of the parameter space we consider, the hierarchy between these couplings may be rather large and, in the absence of further model building, requires a perhaps unnaturally large $B-L$ charge for $\chi$.

\subsubsection{Terrestrial and astrophysical constraints}

For a $U(1)_{B-L}$ gauge boson, the coupling to electrons and protons behaves in much the same way as the dark photon. However, the situation is somewhat different due to the additional coupling of the vector with neutrons, which does not decouple in the $m_{A'} \to 0$ limit (see Appendix~\ref{app:mixing}). For the sun, HB, and RG stars, emission from electrons dominates and the effect of the nucleon coupling is mild, becoming important only for $m_{A'}\lesssim 10^{-2}$ eV~\cite{Hardy:2016kme}.  For SN1987A, the dominant production is nucleon-nucleon scattering. While SN1987A constraints have not been derived for the $U(1)_{B-L}$ case, we can obtain approximate constraints by combining previous results in the literature.  In the  weak coupling regime, we use limits on $U(1)_B$ gauge bosons derived in Ref.~\cite{Rrapaj:2015wgs} and the limits on dark photons from Ref.~\cite{Chang:2016ntp}, whichever is stronger; this approximates the bounds due to the coupling of the $U(1)_{B-L}$ gauge boson with both electrons and nucleons. We derive the result for the trapping limit in a similar way, by combining the trapping due to absorption from   Ref.~\cite{Rrapaj:2015wgs} and the trapping due to decay of $A' \to e^+ e^-$ from Ref.~\cite{Chang:2016ntp}.

\begin{figure}[t]
\centering
\includegraphics[width=0.7\textwidth]{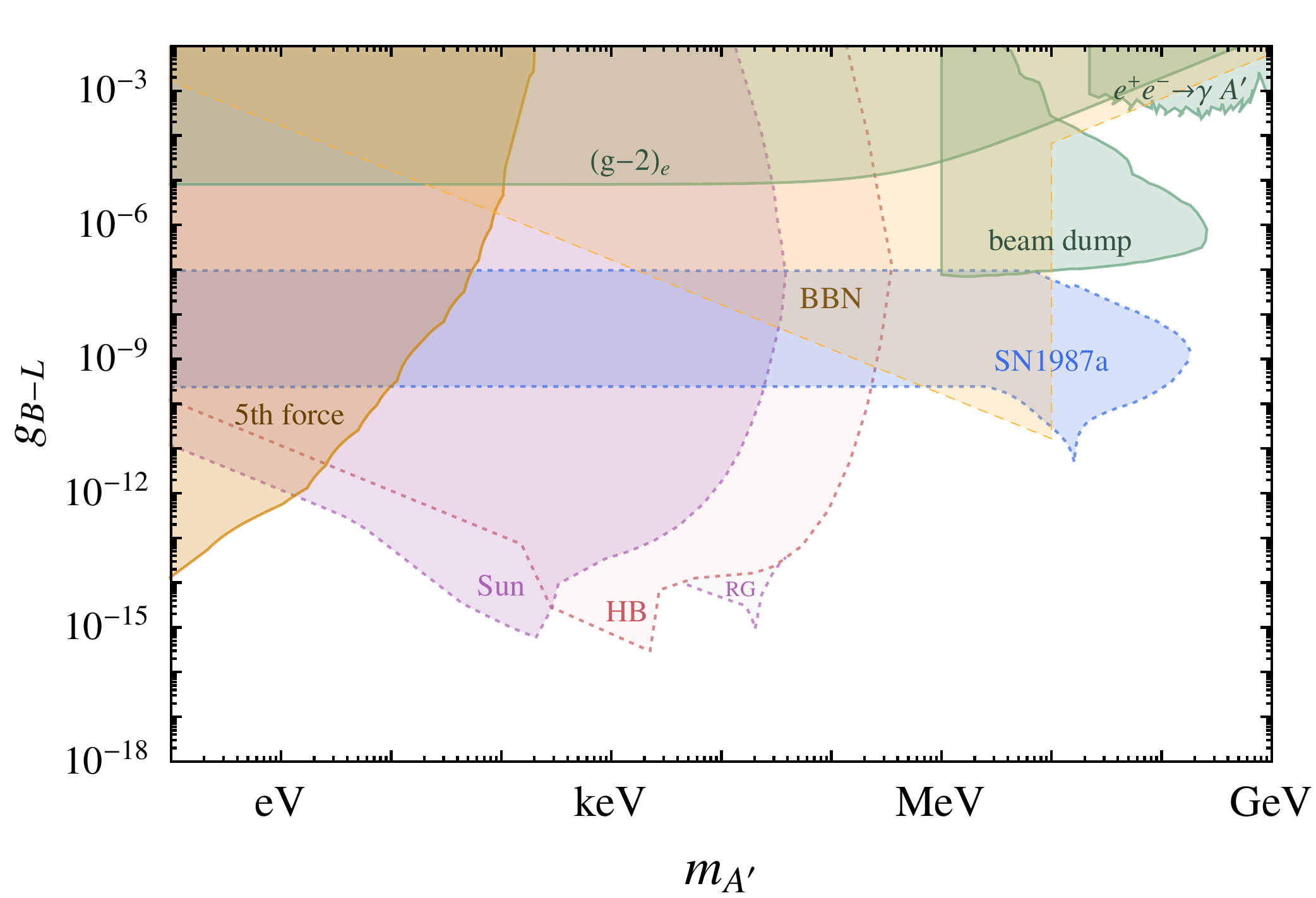} 
\caption{ Constraints on a sub-GeV $U(1)_{B-L}$ mediator. Shown are limits from fifth force searches \cite{Murata:2014nra} and neutron scattering \cite{Leeb:1992qf} (orange), from BBN (yellow), and from stellar cooling in HB stars \cite{Hardy:2016kme} (red), the sun \cite{Hardy:2016kme} (purple), red giants \cite{An:2014twa} (purple), and SN1987A (light blue). Shown in green are also beam dump \cite{PhysRevD.86.095019,Bjorken:2009mm}, BaBar \cite{Lees:2014xha} and $(g-2)_e$ \cite{Davoudiasl:2014kua} constraints.   \label{fig:BLplane}}
\end{figure}

For Majorana neutrinos and $A'$ below the muon threshold, the branching ratio $A'\rightarrow e^+e^-$ is 2/5. We recast the dark photon beam dump \cite{Bjorken:2009mm,PhysRevD.86.095019} and BaBar \cite{Lees:2014xha} constraints to account for the invisible $A'\rightarrow \nu\nu$ decay.  As compared to the leptophilic scalar, the beam dump constraints are truncated below $2m_e$ due to the absence of the $A'\rightarrow\gamma\gamma$ mode. The constraint on $(g-2)_e$ is adapted from Ref.~\cite{Davoudiasl:2014kua}. Similar to the hadrophilic scalar in Section~\ref{sec:nucleon_constraints}, there are fifth force constraints.  The summary of stellar, fifth force, and terrestrial constraints is shown in Fig.~\ref{fig:BLplane}. Note that there are additional laboratory constraints due to neutron scattering, $\nu-e$ scattering, and other atomic physics probes. Since these do not change the conclusions, for clarity we have not shown all of the bounds; for a summary of these limits as well as new bounds derived from isotope shift measurements, see Refs.~\cite{Frugiuele:2016rii,Delaunay:2017dku}.

\begin{figure}[t]
\includegraphics[width=0.49\textwidth]{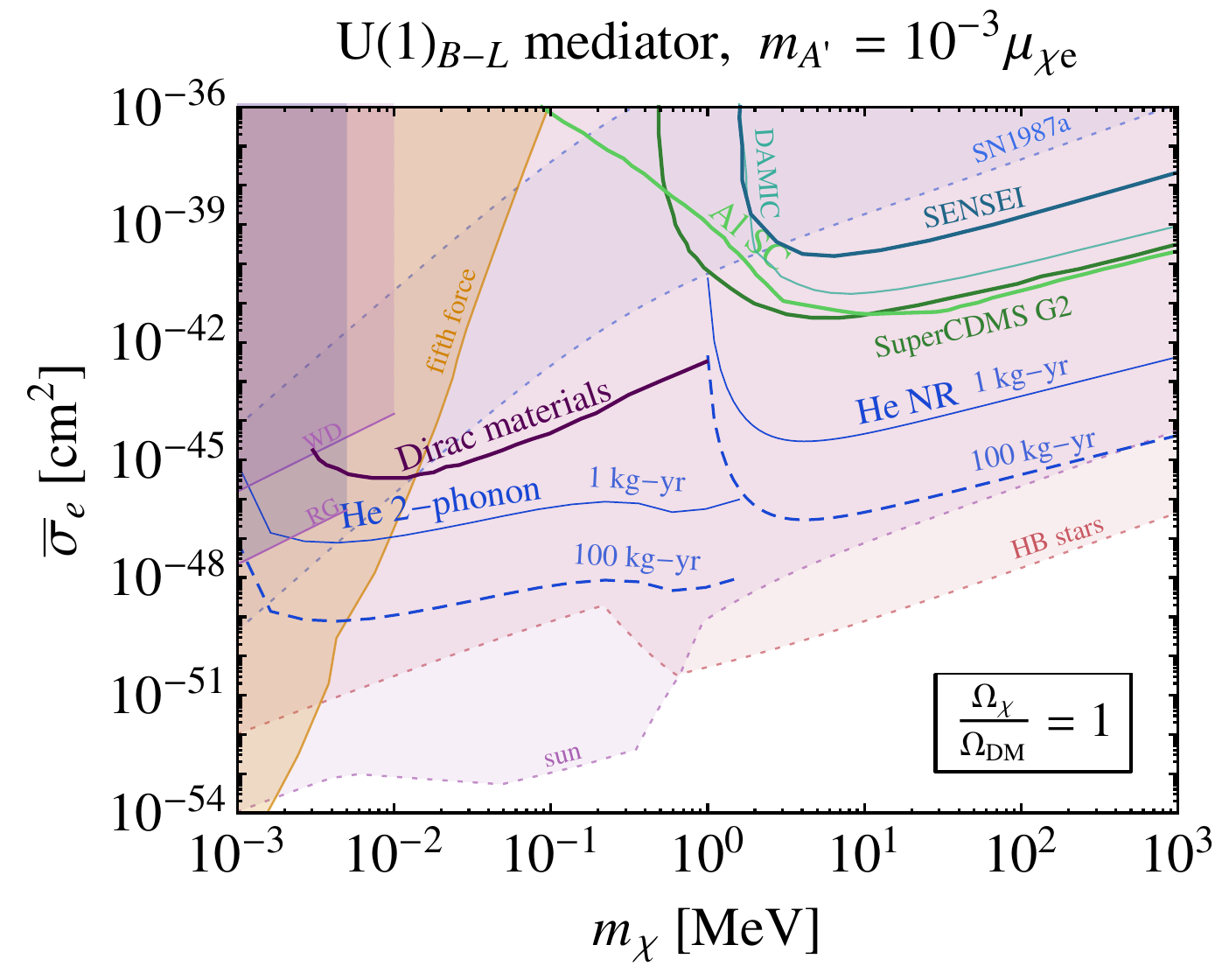}\hfill
\includegraphics[width=0.49\textwidth]{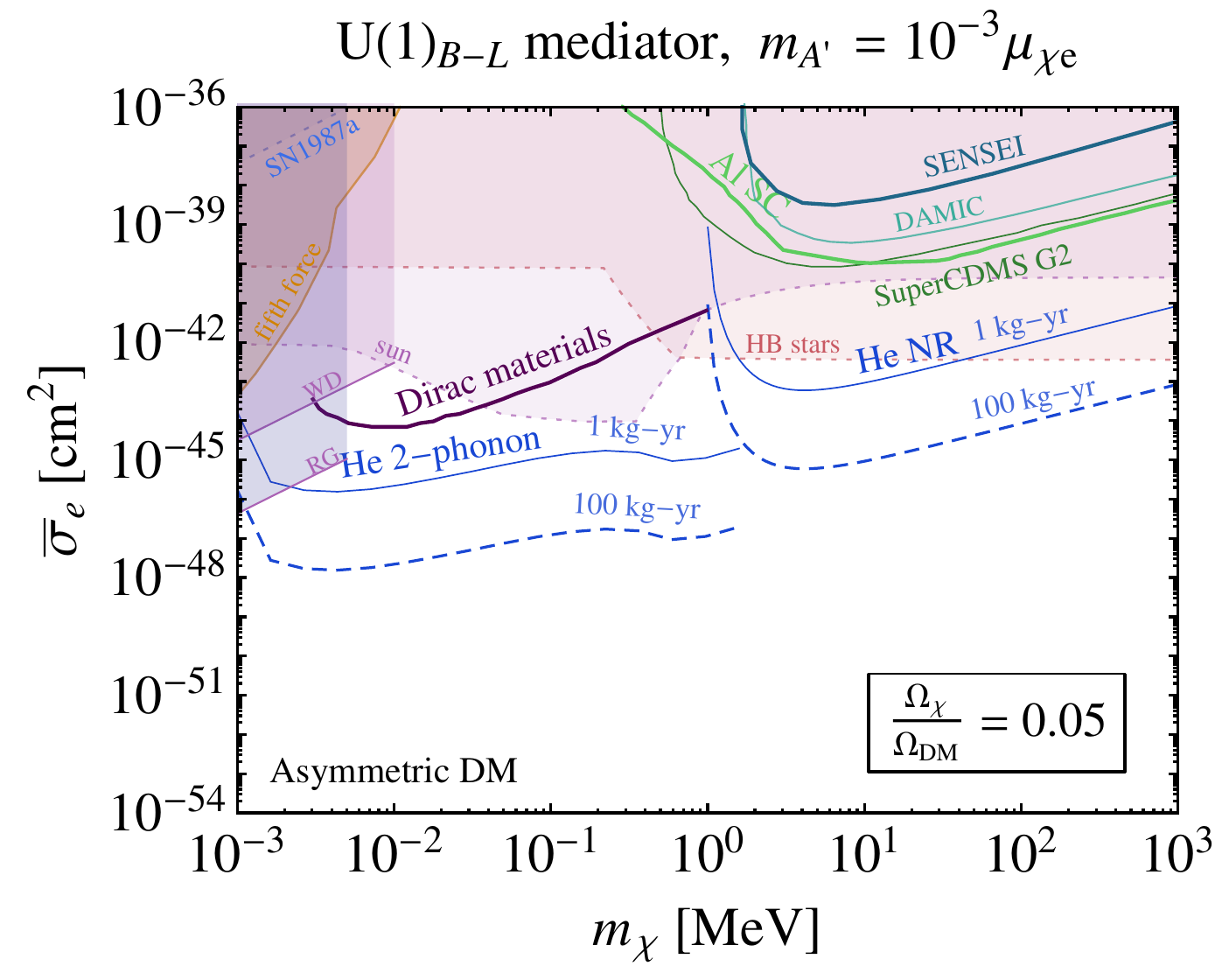}
\caption{   Direct detection cross section vs $m_\chi$ for the case that $\chi$ composes all the dark matter (left), or that $\chi$ is a $5\%$ sub-component of the dark matter (right). We show the projected reach with a kg-year exposure for SuperCDMS G2+, SENSEI-100g, DAMIC-1K~\cite{Battaglieri:2017aum}, superfluid helium~\cite{Knapen:2016cue}, an aluminum superconductor (Al SC) target~\cite{Hochberg:2015fth}, and Dirac materials~\cite{Hochberg:2017wce}.\label{fig:BL_crosssection}
Constraints labeled with ``WD'' and ``RG'' refer to stellar constraints on $\chi$ production itself, as in Fig.~\ref{fig:darkphoton}, while those from the sun, HB stars and SN1987a are bounds  on emission of $A'$, as in Fig.~\ref{fig:BLplane}.}   
\end{figure}

\subsubsection{Cosmology and results}

For Majorana neutrinos, the $A'$ couples to the active neutrinos through the axial current. This implies that the $A'\to \nu\nu$ decay can keep the $A'$ in equilibrium with the neutrinos after the neutrinos decouple from the electron-photon plasma. For $m_{A'}\lesssim 10$ MeV, this is excluded when comparing the deuterium abundance \cite{Cyburt:2015mya,Cooke:2013cba} with the predictions in \cite{Boehm:2013jpa}, similar to what was done in Fig.~\ref{fig:celine} for the leptophilic scalar. For $m_{A'}\gtrsim10$ MeV, the neutrinos can remain in equilibrium with the electron through off-shell $A'$ exchange which would also increase $\Neff$. The dominant process in this case is $e^-\nu\rightarrow e^-\nu$ scattering with a cross section of 
\begin{equation}
\sigma_{e^-\nu\rightarrow e^-\nu}\approx\frac{g^4_{B-L}}{6 \pi}\frac{s}{m_{A'}^4 },
\end{equation}
for which we require that the thermal averaged rate at $T\approx1$ MeV is smaller than the Hubble expansion. The constraints from the decay and scattering processes are indicated by the yellow line in Fig.~\ref{fig:BLplane}. We note that our estimates do not include plasma corrections, and refer to \cite{Heeck:2014zfa} for a treatment of these corrections in the context of a $B-L$ boson with Dirac neutrinos.   

In the heavy mediator regime, where both the dark matter and the mediator thermalize with the SM, $m_{\chi,A'}\gtrsim 10$ MeV is allowed by BBN for reasonably large coupling. Since our focus is primarily on sub-MeV DM, we do not further elaborate here, other than noting that the beam dump constraints Fig.~\ref{fig:BLplane} do not apply if $y_\chi \gg g_{B-L}$ and $2m_\chi<m_{A'}$.
We instead focus on the light mediator regime, for which the allowed parameter space in the $m_\chi$ vs $\bar \sigma_e$ plane is shown in Fig.~\ref{fig:BL_crosssection}, compared to the projected reach for various proposed detectors. Note that the reach for superconductors is significantly weaker than other meV-threshold targets such as superfluid helium and Dirac materials due to in-medium screening effects. If $\chi$ composes all of the dark matter, none of the proposed targets have sensitivity once the SIDM constraints are accounted for. If $\chi$ is a subcomponent of the dark matter, such that the SIDM constraints can be relaxed, then there is some accessible parameter space with superfluid helium or Dirac materials.

\subsubsection{Comments on $U(1)_B$}

As an alternative to gauging $B-L$, one may consider gauging baryon number only. In contrast to $U(1)_{B-L}$, the anomalons needed to render this gauge symmetry anomaly free are not all singlets under the $SU(2)_L\times U(1)_Y$, though it is possible to construct a fully color-neutral set of states (see e.g.~\cite{Dobrescu:2015asa}). Constraints from LEP demand that they are generally heavier than $\gtrsim 100$ GeV. The self-consistency of the low energy effective theory therefore requires $\frac{m_{A'}}{g_B}\gtrsim 100$ GeV, giving a strong bound on $g_B$ for light mediators~\cite{Dobrescu:2014fca}.  A further consequence of the anomalous nature of the symmetry is that certain exotic meson decays involving the longitudinal component of the $A'$ are enhanced \cite{Dror:2017ehi}.

\emph{A priori} one may expect that the constraints on a $U(1)_B$ gauge boson would be greatly relaxed, in particular on the cosmology front since the baryon density drops sharply during the QCD phase transition. However, this is generically not the case, since the $U(1)_B$ always develops a radiative kinetic mixing term with the standard model photon from quark loops, which reintroduces couplings to electrons. Given the rather large number of flavors in the SM, this term is not particularly small, although a mild cancellation exists between the up and down sectors. In particular, the mixing parameter is
\begin{equation}
	\epsilon \approx - \frac{g_B e}{\pi^2}\frac{N_c}{3}  \sum_{f} Q_{f} \int_0^1\!\!dx\; x(1-x)\log\left[\frac{x(1-x)\Lambda_{QCD}^2 + m_f^2}{x(1-x)\Lambda^2 + m_f^2}\right] 
\end{equation}
with $f$ running over all 6 quarks and $Q_f$ and $m_f$ the quark electric charges and masses, respectively. We cut the running off at $\Lambda_{QCD}\approx 1$ GeV, since there are no additional states carrying baryon number below this scale. The parameter $\Lambda$ is the high energy threshold at which $\epsilon\approx 0$. A natural choice for $\Lambda$ is the GUT scale, which yields $\epsilon \approx -0.33\, g_B$, but in principle a scale as low as a few TeV is possible. If we follow the latter (more conservative) avenue and fix $\Lambda\approx 5$ TeV, we find $\epsilon \approx -0.05\, g_B$. It is important to keep in mind that these values are estimates at best, since one expects sizable higher order QCD corrections to this operator. The generic point, however, is that the coupling of a $U(1)_B$ mediator to leptons is not parametrically suppressed to the extent that stellar and BBN constraints from electrons can be neglected, and as such we do not expect a $U(1)_B$ mediator to be a significant exception to arguments presented in this section.

\section{Conclusions\label{sec:conclusion}}

We have considered simplified models for light DM, scattering off nucleons or electrons via scalar or vector mediators in direct detection experiments. 
These models are tightly constrained by stellar cooling arguments and fifth force experiments (in the light mediator limit, for mediator masses below an MeV), or by BBN and CMB constraints on thermalization of the dark sector (in the case of more massive mediators).  The bounds restrict much of the parameter space that can be probed by current proposals for light DM direct detection, especially for dark matter with mass below an MeV.
We highlight the simplified models of sub-MeV DM that satisfy all constraints:
\begin{itemize}
\item DM interacting with nucleons via a mediator heavier than $\sim 100 \mbox{ keV}$, though this is in $\sim 2\sigma$ tension with current BBN bounds on $\Neff$;
\item DM interacting with the standard model via a ultralight kinetically mixed dark photon, and where the DM is populated non-thermally through a mechanism such as freeze-in; 
\item A DM sub-component interacting with nucleons or electrons via a very light scalar or vector mediator.
\end{itemize}
Given the importance of the BBN, CMB, and stellar cooling constraints in our understanding of sub-GeV dark sectors, further exploration of these bounds  and understanding their model-dependence is certainly warranted.

\begin{acknowledgments}
We thank Claudia Frugiuele, Dan Green, Ed Hardy, Rob Lasenby, Zoltan Ligeti, Sam McDermott, Michele Papucci, Surjeet Rajendran, Sean Tulin, Haibo Yu, and Yiming Zhong for useful discussions. We also thank Robert Lasenby for discussions leading to an updated SN1987a bound for the leptophilic model, and Surjeet Rajendran and Dan Green for discussions leading to a numerical correction in the BBN constraints for the B-L model. The authors are supported by the DOE under contract DE-AC02-05CH11231. TL is further supported by NSF grant PHY-1316783.  We thank the Aspen Center for Physics, supported by the NSF Grant No. PHY-1066293, for hospitality while parts of this work were completed.
\end{acknowledgments}

\appendix
\section{Vacuum stability for scalar $\chi$ model \label{app:stability}}

The dark matter interactions in Eq.~\eqref{eq:nucleonmodel} and Eq.\eqref{eq:electronmodel} contain a potential for the scalar $\chi$ and $\phi$,
\begin{align}
	V \supset \frac{1}{2}m_\chi^2\chi^2 + \frac{1}{2}m_\phi^2 \phi^2 + \frac{1}{2}y_\chi m_\chi \phi \chi^2  
\end{align}
which is unbounded from below for finite $\chi$ and $\phi \rightarrow -\infty$. This issue is easily addressed by adding the quartic couplings
\begin{align}
	V \supset  \frac{\lambda_\phi}{4!}\phi^4 + \frac{\lambda_\chi}{4!}\chi^4 +  \frac{\lambda_{\phi\chi}}{4}\phi^2\chi^2.  
\end{align}
Even though the run-away direction is lifted if all quartics are positive, there may still be dangerous false vacua. There is no completely general analytic solution for the positions and energies of the vacua of this potential. However, we can check that there exist self-consistent choices for the quartics such that the origin is the only critical point in the potential. We also require that the quartic couplings do not affect our results on $\chi$ self-interactions and thermalization with the mediator, respectively.

In the light mediator regime, where $m_\phi\ll m_\chi$, the potential has a unique minimum at the origin for $\lambda_\phi\approx 1$, $\lambda_\chi\approx0$ and $\lambda_{\phi\chi} \approx y_\chi$.  Since $\lambda_{\phi\chi}$ only contributes to the self-interaction cross section at loop-level, its contribution can be neglected. $\lambda_{\phi\chi}$ does contribute to the $\phi$-$\chi$ thermalization; however, in the light mediator case there is no bound on this process since the dark sector is not in thermal contact with the standard model below the QCD phase transition.

In the heavy mediator regime for the nucleon coupling model, where $m_\phi\gtrsim m_\chi$, we always require $y_\chi\ll 1$ due to the $\chi-\phi$ thermalization constraint. For example, the following choice suffices to stabilize the potential at the origin for $m_\chi\approx m_\phi$ and  $\lambda_\phi\approx 1$:
\begin{equation}
\begin{array}{ll}
 \lambda_\chi \approx  \lambda_{\phi\chi}\approx 10^{-10}\ , \ \textrm{and}  \ & y_\chi\approx 10^{-4}\\
\end{array}
\end{equation}
 where the choice for $y_\chi$ is representative for the values shown in the blue dashed line in Fig.~\ref{fig:alphaXtherm}. (For smaller $y_\chi$, the vacuum stability condition clearly is easier to satisfy.) Since $\lambda_{\phi\chi},\lambda_{\chi}\ll y^2_\chi$ is possible, the presence of these additional quartics can be neglected in both the thermalization and the self-interaction computations.

For the heavy mediator scenario with electron coupling, we allowed the mediator to thermalize with the dark matter (see Section~\ref{sec:cosmo_electron}). Then $y_\chi$ is bounded by SIDM constraints and can be as large as $\approx 0.1-1$. The following choice is enough to stabilize the potential
\begin{equation}
\begin{array}{lll}
 \lambda_\chi \approx \lambda_{\phi\chi}\approx 10^{-3}\ , \ \textrm{and} \ &  y_\chi\approx 10^{-1},
\end{array}
\end{equation}
again for $m_\chi\gtrsim m_\phi$ and  $\lambda_\phi\approx 1$. Again since $\lambda_{\phi\chi}, \lambda_\chi \ll y_\chi^2$, the corrections to the $\chi$ self-interaction cross section can be neglected for our purposes. 

Finally, the perturbativity constraint requires that the one-loop correction to the $\frac{1}{2}y_\chi m_\chi \phi \chi^2$ coupling is parametrically smaller than the tree-level contribution. In the non-relativistic limit, the one-loop correction is given by
\begin{align}
\frac{y_\chi^3 m_\chi}{16\pi^2} \left[ \log\left( \tfrac{m_\phi}{m_\chi}\right) - \tfrac{m_\phi}{\sqrt{4 m_\chi^2 - m_\phi^2}} \left( \tan^{-1} \tfrac{m_\phi}{\sqrt{4 m_\chi^2 - m_\phi^2}} - \tan^{-1} \tfrac{m_\phi^2 - 2 m_\chi^2}{m_\phi \sqrt{4 m_\chi^2 - m_\phi^2}} \right) \right] 
\end{align}
which we require to be smaller than $y_\chi m_\chi$.
In the $m_\phi\ll m_\chi$ limit, the perturbativity constraint simplifies to 
\begin{equation}
	\frac{y_\chi^2}{16\pi^2}<1 \, ,
\end{equation}
while for $m_\chi=m_\phi$ the result is
\begin{equation}
	\frac{y_\chi^2}{48\sqrt{3}\pi}<1.
\end{equation}
Both cases are consistent with the requirement of $y_\chi\lesssim 4\pi$  from naive dimensional analysis. Throughout this paper we conservatively impose $y_\chi<1$.

\section{Self-interaction cross sections\label{app:SIDM}}

In this appendix, we review scattering of distinguishable dark matter particles. Here $\sigma_T$ is the transfer cross section, defined as the scattering cross section weighted by the momentum transfer,
\begin{equation}
\sigma_T = \int d\Omega \frac{d \sigma}{d \Omega}(1 - \cos \theta).
\end{equation}
In the Born approximation, where $\alpha_\chi m_\chi / m_\phi \ll 1$, the transfer cross section for DM interacting via a Yukawa potential is \cite{Tulin:2017ara}
\begin{equation}
\sigma_T^{\rm born} \approx \frac{8 \pi \alpha_\chi^2}{m_\chi^2 v^4}\left[\log(1+R^2)-R^2/(1+R^2)\right],
\end{equation}
where $R \equiv m_\chi v/m_\phi$.
When the mediator is heavy, such that $R \ll 1$, the coupling constant corresponding to a cross section of $1 \mbox{ cm}^2/\mbox{g}$ is
\begin{equation}
\alpha_\chi \lesssim  0.02 \left(\frac{1 \mbox{ keV}}{m_\chi}\right)^{1/2} \left(\frac{m_\phi}{1 \mbox{ MeV}}\right)^2.
\end{equation}
For  the ultra-light mediator limit where $R \gtrsim 1$, we instead have
\begin{align}
	\alpha_\chi \lesssim 10^{-10} \times  \left(\frac{m_\chi}{1 \mbox{ MeV}}\right)^{3/2},
\end{align}
where we took $v \sim 10^{-3}$.  Note that $R \gtrsim 1$ also corresponds to the classical regime;  in the classical limit, and assuming an attractive potential, analytic formulae for the cross section have been obtained that are valid even for the non-perturbative regime. However, the Born approximation is more accurate for $\alpha_\chi m_\chi/m_\phi \ll 1$, even in the classical regime, and so we use the Born result everywhere.\footnote{We thank Haibo Yu for this comment.} See for example Refs.~\cite{Tulin:2012wi,Tulin:2013teo}. Finally, comparing with the results in Section~\ref{sec:selfint}, we see that the limiting forms of the self-interaction cross section and the resulting bounds on $\alpha_\chi$ are very similar despite the different models considered.

\section{Thermalization of the mediator\label{app:therm}}

In Section~\ref{sec:cosmo}, we estimated when a light scalar mediator interacting with gluons would decouple from the standard model thermal bath. A full calculation of the process $ g g \to \phi g$ requires accounting for thermal gluon masses, which further regulate the $t-$channel collinear divergence. For comparison, here we review some results for light pseudoscalar thermalization with gluons that include finite temperature effects.

We normalize the pseudoscalar-gluon coupling as $\frac{\alpha_s}{4\Lambda} a G^a_{\mu \nu}\tilde G^a_{\mu \nu}$ where $a$ is the pseudoscalar. Ref.~\cite{Salvio:2013iaa} found a pseudoscalar production rate of
\begin{align}
	\gamma_a =  \frac{\alpha_s^2  \zeta(3) T^6}{\pi^3 \Lambda^2} F_3 (m_3/T)  \simeq  \frac{4 \alpha_s^2 \zeta(3) T^6}{(\pi)^3 \Lambda^2} 
\end{align}
where $\gamma_a  \sim n_g^2 \sigma v$ is the collision term in the Boltzmann equation and $n_g$ is the thermal gluon density.
Comparing $\gamma_\phi/n_g$ with the Hubble expansion, this gives $\Lambda \lesssim 3 \times 10^9\ \GeV \sqrt{T/\GeV}$ for $a$ production to be out of equilibrium at temperature $T$.
Similarly, an earlier result using the Hard Thermal Loop approximation found~\cite{Graf:2010tv}
\begin{align}
	\gamma_a =  \frac{ 4 \zeta(3) \alpha_s T^6}{\pi^2 \Lambda^2} \times \left( \ln \frac{T^2}{m_g^2} + 0.4 \right),
\end{align}
resulting in a similar condition on $\Lambda$. 

Assuming that the results above can also be applied to a scalar coupling with gluons, we find the following condition on the scalar nucleon coupling:
\begin{align}
	y_n \lesssim \textrm{few} \times 10^{-10}
\end{align}
such that $\phi$ decouples from the thermal bath before the QCD phase transition. In the main text we simply use $y_n \lesssim 10^{-9}$ as an approximate condition.

\section{In-medium couplings of light vectors \label{app:mixing}}

We begin by writing the vacuum Lagrangian in the basis where the kinetic-mixing term has been rotated away, such that our discussion can be applied for the dark photon and for $B-L$. Consider the {\it vacuum} Lagrangian
\begin{align}
	{\cal L}_{\rm vac} = & -\frac{1}{4} F_{\mu \nu} F^{\mu \nu} - \frac{1}{4}  F'_{\mu \nu} F'^{\mu \nu} + \frac{m_{A'}^2}{2} {A'}^\mu {A'}_\mu \\ \nonumber  
& + J_{\rm EM}^\mu \left(e A_\mu + g {A'}_\mu\right) + g J_{{\rm SM}' \, }^\mu {A^\prime}_\mu  + g_{\rm DM} J_{\rm DM}^\mu {A'}_\mu.
\end{align}
Here we have separated the coupling of the new $A'$ gauge boson to SM particles into two pieces, the coupling to the EM current and everything else (accounted for in $J_{\rm SM'}$). Additionally, we include the coupling of $A'$ with dark matter, which we write as $g_{\rm DM}$ to account for a possible large hierarchy between the $A'$ couplings with the SM and the dark matter.

Taking a coupling hierarchy $e \gg g$, in-medium effects generate mass terms
\bea
{\cal L}_{\rm IM-mass} = \frac{m_A^2}{2} A^\mu A_\mu + \epsilon m_A^2 A^\mu {A'}_\mu,
\eea
where $\epsilon = g/e$ and $m_A$ is the in-medium mass of the photon due to the charged particle density.
Now we rotate to the mass basis by the choice
\begin{align}
	A^\mu = \tilde A^\mu + \frac{\epsilon m_{A}^2}{m_{A'}^2 - m_{A}^2}  \tilde A'^\mu,
~~~~~A'^\mu = \tilde A'^\mu - \frac{\epsilon m_{A}^2}{m_{A'}^2 - m_{A}^2}  \tilde A^\mu,
\end{align}
such that the mass mixing is eliminated up to terms of order $O(\epsilon^2)$. Then the total in-medium Lagrangian, dropping the $O(\epsilon^2)$ terms, becomes
\begin{align}
	{\cal L}_{\rm IM} =&  -\frac{1}{4} \tilde F_{\mu \nu} \tilde F^{\mu \nu} 
			- \frac{1}{4} \tilde F'_{\mu \nu} \tilde F'^{\mu \nu} 
			+ \frac{m_{A}^2}{2} \tilde A^\mu \tilde A_\mu
			+ \frac{m_{A'}^2}{2} \tilde A'^\mu \tilde {A'}_\mu 
			+ g J_{{\rm SM}' \, }^\mu \tilde {A^\prime}_\mu  \nonumber  \\
	 &+ J_{\rm EM}^\mu \left(e \tilde A_\mu + \frac{e \epsilon \, m_{A'}^2}{m_{A'}^2 - m_{A}^2} \tilde {A'}_\mu\right)    + g_{\rm DM} J_{\rm DM}^\mu \left(  \tilde {A^\prime}_\mu - \frac{\epsilon m_{A}^2}{m_{A'}^2 - m_{A}^2}  \tilde A_\mu \right).
	\label{eq:inmedium_L}
\end{align}
Here we kept the terms of $O(g_{\rm DM} \epsilon)$ since it is possible that $g_{\rm DM} \gg g$.

For the piece of the $\tilde A'$ coupling that is proportional to the EM current, the mass mixing implies that  the effective in-medium coupling will decouple in the limit $m_{A'} \ll m_A$. In stars such as the sun, $m_A$ can be of order keV, while $m_A \sim 20\, \MeV$ for SN1987A, leading a significant effect for low mass vectors. 
This is familiar in the case of the dark photon where $J_{\rm SM'} = 0$ and taking $\epsilon$ as the kinetic mixing parameter. Here it is known that the in-medium kinetic mixing parameter is suppressed by $\approx m_{A'}^2/m_A^2$ in the small $m_{A'}$ limit. Due to a relative enhancement by $\sim E/{m_{A'}}$ for production of the longitudinal modes, the resulting dominant stellar emission scales as $\epsilon m_{A'}$ in the low mass limit~\cite{An:2013yfc,An:2013yua,An:2014twa}.

For $B-L$ the same analysis applies, however there is also an interaction of the $\tilde A'$ with neutrons, included in $J_{\rm SM'}$. As shown above, this coupling does {\it not} decouple in the low mass limit, leading to stellar constraints that are independent of $m_{A'}$. For the sun, HB stars, and RG stars, production off of neutrons is a smaller contribution than production off of charged particles, so this effect is only relevant for $m_{A'}$ well below an eV~\cite{Hardy:2016kme}. However, in a supernova the emission of $A'$ in nucleon-nucleon scattering is important. As described in Section~\ref{sec:B_L}, we thus obtain SN1987A bounds on $B-L$ by combining limits on dark photons from Ref.~\cite{Chang:2016ntp} and limits on $U(1)_B$ from Ref.~\cite{Rrapaj:2015wgs}, whichever gives the larger effect for a given $m_{A'}, g$.

The interactions above also allow production of the DM in the star, via an off-shell $\tilde A'$ or off-shell $\tilde A$. Although the coupling of $\tilde A'$ to SM charged particles is suppressed for $m_{A'} \ll m_A$, from Eq.~\eqref{eq:inmedium_L} the induced coupling of $\tilde A$ with dark matter is given by $\epsilon g_{DM}$ in this limit. If $g_{DM}$ is relatively large, then production of DM furnishes another form of stellar constraint, and we can apply limits derived for the millicharged DM case taking $Q_{DM} \approx (\epsilon g_{DM}/e) $. This can be directly translated to an upper bound on the direct detection cross section  (in addition to the constraints on $\epsilon$ and $g_{DM}$ from stellar emission and self-interactions, respectively).

\bibliographystyle{jhep}
\bibliography{lightdm}

\end{document}